\documentclass[12pt]{article}
\usepackage{amsmath,amssymb}
\usepackage{bm}

\catcode`\@=11

\ifx\documentclass\undefined
 \def\@sect#1#2#3#4#5#6[#7]#8{\ifnum #2>\c@secnumdepth
     \let\@svsec\@empty\else
     \refstepcounter{#1}\edef\@svsec{\csname prefix#1\endcsname
        \csname the#1\endcsname\hskip 1em}\fi
     \@tempskipa #5\relax
      \ifdim \@tempskipa>\z@
        \begingroup #6\relax
          \@hangfrom{\hskip #3\relax\@svsec}{\interlinepenalty \@M #8\par}%
        \endgroup
       \csname #1mark\endcsname{#7}\addcontentsline
         {toc}{#1}{\ifnum #2>\c@secnumdepth \else
                      \protect\numberline{\csname the#1\endcsname}\fi
                    #7}\else
        \def\@svsechd{#6\hskip #3\relax  %% \relax added 2 May 90
                   \@svsec #8\csname #1mark\endcsname
                      {#7}\addcontentsline
                           {toc}{#1}{\ifnum #2>\c@secnumdepth \else
                             \protect\numberline{\csname the#1\endcsname}\fi
                       #7}}\fi
     \@xsect{#5}}
\else
    \def\@seccntformat#1{\csname prefix#1\endcsname
        \csname the#1\endcsname\quad}
\fi

\@addtoreset{equation}{section}   % Makes \section reset 'equation' counter.
\def\theequation{\arabic{section}.\arabic{equation}}

\def\thebibliography#1{\section*{References\@mkboth
 {REFERENCES}{REFERENCES}}\list
 {\leftbibmark\arabic{enumi}\rightbibmark}{
 \settowidth\labelwidth{\leftbibmark #1\rightbibmark}\leftmargin\labelwidth
 \advance\leftmargin\labelsep
 \usecounter{enumi}}
 \def\newblock{\hskip .11em plus .33em minus -.07em}
 \sloppy\clubpenalty4000\widowpenalty4000
 \sfcode`\.=1000\relax}

\def\@citex[#1]#2{\if@filesw\immediate\write\@auxout{\string\citation{#2}}\fi
  \def\@citea{}\@cite{\@for\@citeb:=#2\do
    {\@citea\def\@citea{,\penalty\@m\ }\@ifundefined
       {b@\@citeb}{{\bf ?}\@warning
       {Citation `\@citeb' on page \thepage \space undefined}}%
\hbox{\csname b@\@citeb\endcsname\citemarkdelim}}}{#1}}

\def\@cite#1#2{\leftcitemark{#1 \if@tempswa , #2\fi}\rightcitemark}
\def\leftcitemark{[}
\def\rightcitemark{]}
\def\citemarkdelim{}
\def\leftbibmark{[}
\def\rightbibmark{]}

\catcode`\@=12

\usepackage{amsmath}
\usepackage{amssymb}

\begin{document}

\begin{titlepage}

\vspace{2cm}

\begin{center}\large\bf
Manifestly gauge invariant theory \\ 
of the nonlinear cosmological perturbations \\
in the leading order of the gradient expansion 
\end{center}

\begin{center}
Takashi Hamazaki\footnote{email address: yj4t-hmzk@asahi-net.or.jp}
\end{center}

\begin{center}\it
%\begin{center}
Kamiyugi 3-3-4-606 Hachioji-city\\
Tokyo 192-0373 Japan\\
\end{center}

%T2>Abstract
\begin{center}\bf Abstract\end{center}

In the full nonlinear cosmological perturbation theory in the 
leading order of the gradient expansion, all the types of the 
gauge invariant perturbation variables are defined.
The metric junction conditions across the spacelike transition
hypersurface are formulated in a manifestly gauge invariant manner.
It is manifestly shown that all the physical laws such as the 
evolution equations, the constraint equations, and the junction conditions
can be written using the gauge invariant variables which we defined
only.
Based on the existence of the universal adiabatic growing mode in 
the nonlinear perturbation theory and the $\rho$ philosophy where
the physical evolution are described using the energy density $\rho$ 
as the evolution parameter,
we give the definitions of the adiabatic perturbation variable and 
the entropic perturbation variables in the full nonlinear perturbation
theory.
In order to give the analytic order estimate of the nonlinear parameter
$f_{NL}$, we present the exponent evaluation method.
As the models where $f_{NL}$ changes continuously and becomes large,
using the $\rho$ philosophy, we investigate the non-Gaussianity induced by 
the entropic perturbation of the component which does not govern the 
cosmic energy density, and we show that in order to obtain the significant 
non-Gaussianity it is necessary that the scalar 
field which supports the entropic perturbation is extremely small 
compared with the scalar field which supports the adiabatic perturbation.

\vspace{1cm}

  
PACS number(s):98.80.Cq

\end{titlepage}

\section{Introduction and summary}

In the inflationary scenario, the quantum fluctuations of the scalar fields
driving the inflationary expansion of the universe are the origins, that is 
the seed perturbations of the temperature fluctuations of the cosmic microwave 
background radiation (CMB) and the cosmic large scale structures such as galaxies 
and clusters of galaxies.
These seed perturbations generated in the horizon during the inflationary 
expansion are stretched and go out of the horizon.
They stay outside the horizon until they return into the horizon in the 
Friedman expansion stage.
Therefore in order to compare the theory with the observation, it is necessary
to solve the evolutions of the cosmological perturbations on superhorizon scales
using the concrete theoretical models such as the various inflation and the 
reheating scenarios. \cite{Kodama1996} \cite{Hamazaki1996} \cite{Kodama1998} 
\cite{Gordon2001} \cite{Hamazaki2002} \cite{Hamazaki2004} \cite{Hamazaki2008}
\cite{Hamazaki2008.2}
Fortunately as for the evolutions of the cosmological perturbations in the 
long wavelength limit, the exact solution is constructed in terms of the
evolution of the corresponding locally homogeneous universe;
as for the linear perturbations in the papers \cite{Taruya1998} \cite{Kodama1998} 
\cite{Sasaki1998} \cite{Hamazaki2008} and as for the full nonlinear perturbations 
in the papers \cite{Rigopoulos2003} \cite{Lyth2005} \cite{Hamazaki2008.2}.
The final form of the exact solutions of the evolutions of the cosmological 
perturbations in the long wavelength limit is established by  
the Kodama Hamazaki construction (KH construction), as for the linear perturbations
in the papers \cite{Kodama1998} \cite{Hamazaki2008}, and as for the full nonlinear 
perturbations in the paper \cite{Hamazaki2008.2}.
In the KH construction, the physical quantities related with the exactly homogeneous
universe, such as the scalar quantity perturbations, are given as the solutions 
of the evolution equations of the corresponding locally homogeneous universe and
the physical quantities not related with the exactly homogeneous universe,
such as the vector quantity perturbations, are given by solving the first order
evolution equations, that is, the spatial components of the Einstein equations.
It was shown that the second order evolution equations of the spatial unimodular metric 
including the information of the adiabatic decaying mode is exactly solvable.    
In the present paper, we use the KH construction.

The general theory of relativity is a gauge theory.
When we solve the equations of the general theory of relativity, the nondynamical
gauge modes are contained in the solutions.
Therefore in order to extract the dynamical modes only, it is desirable to write
down the equations in terms of the gauge invariant variables only.
In the linear perturbation theory, the program of the gauge invariant perturbation
variables was first performed in the paper \cite{Bardeen1980}, and was
extended so that we can treat the multicomponent systems \cite{Kodama1984}
\cite{Kodama1987} \cite{Mukhanov1992}.
In the second order perturbation theory, the gauge invariant perturbation 
theory was constructed in the papers \cite{Malik2004} \cite{Malik2005} 
\cite{Nakamura2006}.
In our previous paper \cite{Hamazaki2008.2}, the full nonlinear perturbation theory
in the leading order of the gradient expansion was constructed and several main
definitions of the gauge invariant perturbation variables including the nonlinear
Bardeen parameters \cite{Bardeen1980} \cite{Mukhanov1992} \cite{Wands2000} 
\cite{Lyth2005} \cite{Hamazaki2008.2} were presented. 
In the present paper, in a more general way, definitions of all the types of 
the gauge invariant perturbation variables are constructed and it is manifestly shown 
that all the perturbation equations of the physical laws such as the evolution equations, 
the constraint equations and the metric junction conditions can be written by using  
the gauge invariant perturbation variables which we defined only.
By solving the equations of the gauge invariant formulation of the full nonlinear 
perturbations in the leading order of the gradient expansion which we formulated, 
we can extract the full nonlinear physically meaningful, dynamical information of the 
cosmological perturbations on superhorizon scales.

In order to interpret the physics of the evolutionary behaviors of the 
cosmological perturbations, the Adiabatic/Entropic decomposition
(A/E decomposition) of the cosmological perturbations is efficient.
The essence of the A/E decomposition is in defining the adiabatic perturbation
variable and the entropic perturbation variables.
Although the linear version of the A/E decomposition has been already established
\cite{Kodama1984} \cite{Kodama1987}, the satisfactory definitions of the adiabatic 
perturbation variable and the entropic perturbation variables in the nonlinear perturbation 
theory has not been completed yet.
In the present paper, we give the definitions of the adiabatic perturbation variable 
and the entropic pertubation variables which can be used in the nonlinear perturbation theory
by using the fact that the universal adiabatic growing mode always exists in the solutions
in the nonlinear perturbation theory in the long wavelength limit. \cite{Kodama1998}
That is, we call the perturbation variable which does not vanish for the universal 
adiabatic growing mode the adiabatic perturbation variable and we call the perturbation 
variable which vanishes for the universal adiabatic growing mode the entropic perturbation 
variable.
In particular, the adiabatic/entropic perturbation variables which are defined under the 
$\rho$ philosophy where the evolutions of the system are traced by choosing 
the energy density $\rho$ as the evolution parameter, have desirable properties.
All the perturbation variables in this set are continuous across the metric junction 
hypersurface which is defined by $\rho = {\rm const}$ such as the slow rolling-oscillatory
transitions of the scalar fields and the reheating transitions.
The evolution equations of the perturbation variables in this set which can be derived by 
quite easy calculation have very simple expression.
The adiabatic perturbation variable in this set is the well-known Bardeen parameter. 
\cite{Bardeen1980} \cite{Mukhanov1992} \cite{Wands2000} \cite{Lyth2005} \cite{Hamazaki2008.2}

In the near future, more precise observations of CMB will be performed and it is expected
that the information of the nonlinearity of the CMB fluctuations will be obtained.
Motivated by the observational advancement, the models which generate the significant 
nonlinearity characterized by the large non-Gaussianity parameter $f_{NL}$ 
\cite{Komatsu2001} have been proposed; the inhomogeneous end of the inflation 
\cite{Lyth2005.2} \cite{Sasaki2008}, the modulated reheating \cite{Dvali2004},
the curvaton scenario \cite{Ichikawa2008}, the vacuum dominated inflation 
\cite{Alabidi2006} \cite{Byrnes2009}.
The former two cases are related with the metric junction hypersurface which cannot be
defined by $\rho = {\rm const}$ and the large non-Gaussianities $f_{NL}$ are
generated discontinuously on the transition hypersurface.
In the latter two cases, the non-Gaussianity $f_{NL}$ grows continuously and
becomes very large transiently.
In the present paper, we present the exponent evaluation method which enables us
to give the analytic order estimates of the non-Gaussianities $f_{NL}$ in these 
models.
We discuss that the mechanisms which generate the large non-Gaussianities $f_{NL}$
in the latter two cases are common, although in the first case in the latter two 
cases the cosmological term does not exist while in the second case in the latter
two cases the cosmological term exists.
In the latter two cases, the entropic perturbation of the component which does not
govern the cosmic energy density can trigger the growth of the Bardeen parameter
$\zeta_n (\rho)$ and the non-Gaussianity $f_{NL}$, when the scalar fields which 
support the entropic perturbation are very small, since the influences of these 
small scalar fields on the Bardeen parameter $\zeta_n (\rho)$ can become large.

The rest of the present paper is organized as follows.
In the section $2$, we give the definitions of all the types of 
the gauge invariant perturbation variables and show manifestly that
in the long wavelength limit all the perturbation equations of all the physical 
laws derived by the general theory of relativity can be written in the 
gauge invariant manner.
In the section $3$, under the $\rho$ philosophy, we complete the A/E decomposition 
of the full nonlinear perturbations by giving the definitions of the adiabatic 
perturbation variable and the entropic pertubation variables.
In the section $4$, as the application of the A/E decomposition
based on the $\rho$ philosophy formulated in the previous section,
we investigate the evolutions of the cosmological perturbations in the 
universe where the growth of the adiabatic perturbation variable called 
the Bardeen parameter \cite{Bardeen1980} \cite{Mukhanov1992} \cite{Wands2000} 
\cite{Lyth2005} \cite{Hamazaki2008.2} is induced by the entropic perturbation 
of the subdominant component.
We evaluate the non-Gaussianity parameter $f_{NL}$ by the exponent evaluation method. 
We present the condition for which the non-Gaussianity $f_{NL}$ becomes large 
in the models where $f_{NL}$ changes continuously.

\section{the manifestly gauge invariant formulation of the nonlinear cosmological
perturbation theory in the leading order of the gradient expansion}

\subsection{the evolution equations and the constraint equations}

We consider the Einstein equations $G_{\mu \nu} = \kappa^2 T_{\mu \nu}$
where $\kappa^2$ is expressed in terms of the Newtonian gravitational constant
$G$ as $\kappa^2 = 8 \pi G$, using the $3 + 1$ decomposition.  
\cite{Shibata1999} \cite{Tanaka2007} \cite{Hamazaki2008.2}
The Greek indices $\mu, \nu, \cdot \cdot \cdot$ run from $0$ to $3$ and the 
Latin indices $i, j, \cdot \cdot \cdot$ run from $1$ to $3$.
The metric tensor $g_{\mu \nu}$ is expressed as
\begin{eqnarray}
 g_{0 0} &=& - \alpha^2 + \beta_k \beta^k,\\ 
 g_{0 i} &=& \beta_i,\\
 g_{i j} &=& \gamma_{i j},
\end{eqnarray}
where $\alpha$ is the lapse and $\beta_i$ is the shift vector.
The index of $\beta_i$ is raised by $\gamma^{i j}$ which is the inverse matrix 
of $\gamma_{i j}$.
The spatial metric $\gamma_{i j}$ is factorized as
\begin{equation}
 \gamma_{ij} = a^2 \tilde{\gamma}_{ij},
\end{equation}
where $\tilde{\gamma}_{ij}$ is the unimodular matrix whose inverse matrix is 
expressed as $\tilde{\gamma}^{ij}$ and $a$ is the scale factor. 
The energy momentum tensor of the total system $T_{\mu \nu}$ is expressed as
\begin{equation}
 T_{\mu \nu} = (\rho + P) u_{\mu} u_{\nu} + P g_{\mu \nu},
 \label{totalenergymomentum1}
\end{equation}
where $\rho$, $P$ and $u_{\mu}$ are the energy density, the pressure and the four
velocity vector of the total system, respectively.
Because of the normalization condition $u^{\mu} u_{\mu} = -1$, $u_{\mu}$ can be 
parametrized as
\begin{eqnarray}
 u_0 &=& - u^0 \{ \alpha^2 - \beta^k (\beta_k + v_k) \},
 \label{totalenergymomentum2}\\
 u_k &=& u^0 (\beta_k + v_k),
 \label{totalenergymomentum3}
\end{eqnarray}
where $u^0 = g^{0 \mu} u_{\mu}$ is given by
\begin{equation}
 u^0 = \{ \alpha^2 - (\beta_k + v_k) (\beta^k + v^k) \}^{-1/2},
 \label{totalenergymomentum4}
\end{equation}
$v_k$ is the three velocity of the total system and the index of $v_k$ is raised
by $\gamma^{ij}$.
$T_{\mu \nu}$ is expressed as 
\begin{equation}
 T_{\mu \nu} = \sum_{\alpha} T_{\alpha \mu \nu} + T_{S \mu \nu},
\end{equation}
where $T_{\alpha \mu \nu}$ is the energy momentum tensor of the perfect fluid component
$\alpha$ and $T_{S \mu \nu}$ is the energy momentum tensor of all the scalar fields.
$T_{\alpha \mu \nu}$ is expressed by
(\ref{totalenergymomentum1}) (\ref{totalenergymomentum2}) (\ref{totalenergymomentum3})
(\ref{totalenergymomentum4}) where $\rho$, $P$, $u_{\mu}$ and $v_i$ are replaced with
$\rho_{\alpha}$, $P_{\alpha}$, $u_{\alpha \mu}$ and $v_{\alpha i}$, respectively.
$T_{S \mu \nu}$ is expressed by
\begin{equation}
 T_{S \mu \nu} = \sum_a \partial_{\mu} \phi_a \partial_{\nu} \phi_a 
 - \frac{1}{2}
 \left\{ \sum_a g^{\rho \sigma} \partial_{\rho} \phi_a \partial_{\sigma} \phi_a + 2 U \right\}
 g_{\mu \nu}.
\end{equation}
As for the scalar fields, since we cannot decide to which component $a$ each term of the potential
$U$ belongs, $T_{S \mu \nu}$ cannot be decomposed into $T_{a \mu \nu}$.
In this way, the indices of the component $A$ are divided into the perfect fluid indices $\alpha$
and the scalar field indices $a$.
The energy momentum transfer vectors of the perfect fluid component $\alpha$ and the scalar field
component $a$ are expressed by
\begin{eqnarray}
 Q_{\alpha \mu} &=& Q_{\alpha} u_{\mu} + f_{\alpha \mu}, \quad \quad \quad
 u^{\mu} f_{\alpha \mu} =0, \\
 Q_{a \mu} &=& S_a \partial_{\mu} \phi_a,
\end{eqnarray}
where $Q_{\alpha}$ and $f_{\alpha \mu}$ are the energy transfer and the momentum transfer
of the perfect fluid component $\alpha$, respectively and $S_a$ is the source function of the 
scalar field component $a$. 
The energy momentum conservation gives
\begin{equation}
 \sum_{\alpha} Q_{\alpha \mu} + \sum_a Q_{a \mu} =0.
\end{equation}
As for the perfect fluid component $\alpha$, $\nabla_{\mu} T^{\mu}_{\alpha \nu} = Q_{\alpha \nu}$
gives the equation of motion of the perfect fluid component $\alpha$.
As for the scalar fields,  $\nabla_{\mu} T^{\mu}_{S \nu} - \sum_a Q_{a \nu} =0$ can be expressed 
as the linear combination of $\partial_{\nu} \phi_a$.
By assuming that the each coefficient of $\partial_{\nu} \phi_a$ is separately vanishing, we can 
derive the phenomenological equation of motion of the scalar field $\phi_a$,
$\Box \phi_a - \partial U / \partial \phi_a = S_a$.

Since we want to treat the cosmological perturbations on superhorizon scales, we put the gradient
expansion assumptions by using the small parameter $\epsilon$ characterizing the inverse of the 
long wavelength of the cosmological perturbations.
Since the spatial scale of the inhomogeneity of all the physical quantities is of the order of
$1 / \epsilon$, we assign $\partial_i = O(\epsilon)$.
As for the metric, we assign $g_{0 i} = O(\epsilon)$.
For arbitrary vector fields $V_{\mu}$ satisfying $V^{\mu} V_{\mu} = O(1)$ including $u_{\mu}$,
$u_{\alpha \mu}$, we assume that $V_i = O(\epsilon)$.
Therefore $\beta_i$, $\beta^i$, $v_i$, $v^i$, $v_{\alpha i}$, $v^i_{\alpha}$ and $f_{\alpha i}$ 
are of the order of $\epsilon$.
As for the velocity vector of the total system and the perfect fluid component $\alpha$,
the leading order of the gradient expansion can be expressed by
\begin{eqnarray}
 u_0 &=& - \alpha + O(\epsilon^2),\\
 u_i &=& \frac{1}{\alpha} (v_i + \beta_i) + O(\epsilon^3),\\
 u_{\alpha 0} &=& - \alpha + O(\epsilon^2),\\
 u_{\alpha i} &=& \frac{1}{\alpha} (v_{\alpha i} + \beta_i) + O(\epsilon^3).
\end{eqnarray}
As for the momentum transfer vector of the perfect fluid component $\alpha$, 
$u^{\mu} f_{\alpha \mu}=0$ gives
\begin{equation}
 f_{\alpha 0}= 0 + O(\epsilon^2).
\end{equation}

We consider the gauge transformation laws of all the physical quantities.
The gauge transformation laws are written in terms of the Lie derivative.
The Lie derivatives of the quantity with upper index and  the quantity 
with lower index are expressed by
\begin{eqnarray}
 L(T) X^{\mu} &=& T^{\rho} \partial_{\rho} X^{\mu} 
  - \partial_{\rho} T^{\mu}  X^{\rho},\\
  L(T) X_{\mu} &=& T^{\rho} \partial_{\rho} X_{\mu} 
  + \partial_{\mu} T^{\rho}  X_{\rho}.
\end{eqnarray}
The Lie derivative of the tensor field of an arbitrary rank is given by the 
above two definitions and the Leibniz rule.
Because of the gradient expansion assumption, the infinitesimal coordinate transformation
generating the Lie derivative $T^{\rho} \partial_{\rho}$ satisfies $T^i = O(\epsilon)$.
Under the gradient expansion assumption, the Lie derivative of the scalar $S$
is given by
\begin{equation}
 L(T) S = T^0 \dot{S} + O(\epsilon^2),
 \label{scalarlikeobject}
\end{equation}
and the Lie derivative of the vector $V_{\mu}$ is given by
\begin{eqnarray}
 L(T) V_0 &=& T^0 \dot{V}_0 
  + \dot{T}^0  V_0 + O(\epsilon^2),
\label{vectorlikeobject1}\\
 L(T) V_i &=& T^0 \dot{V}_i 
  + \partial_i T^0  V_0 + O(\epsilon^3).
\label{vectorlikeobject2}
\end{eqnarray}
Under the gradient expansion scheme, it is possible that the quantity which is 
not a scalar, for example $\gamma_{ij}$, has the Lie derivative of the scalar type
(\ref{scalarlikeobject}).
So we expand the definitions of the scalar field and the vector field as follows. 
\paragraph{Definition}
The physical quantity which has the Lie derivative (\ref{scalarlikeobject}) 
is called the scalar like object.
The physical quantity which has the Lie derivative (\ref{vectorlikeobject1}) 
(\ref{vectorlikeobject2}) is called the vector like object.

Following these definitions, the physical quantities such as $a$, $\gamma_{ij}$,
$\tilde{\gamma}_{ij}$, $\gamma^{ij}$, $\tilde{\gamma}^{ij}$ are the scalar like 
objects and the $\partial_{\mu}$ derivative of these quantities is the vector
like object.
We can demonstrate the following propositions easily.
\paragraph{Proposition $1$}
For a scalar like object $S$, $\partial_{\mu} S$ is a vector like object.

Please use $L(T) \partial_{\mu} A = \partial_{\mu} \{ L(T)A \}$ for an arbitrary
quantity $A$.
\paragraph{Proposition $2$}
For two arbitrary vector like objects $A_{\mu}$, $B_{\mu}$,
\begin{equation}
 \frac{A_0}{B_0}, \quad A_i - \frac{A_0}{B_0} B_i
\end{equation}
are scalar like objects.
\paragraph{Corollary}
For a scalar like object $A$, $D_t A$ where
\begin{equation}
 D_t := \frac{1}{\alpha} \frac{\partial}{\partial t}
\end{equation}
is also a scalar like object.

Please notice that $u_0 = - \alpha + O(\epsilon^2)$, and that $\partial_{\mu} A$
is a vector like object.  
\paragraph{Corollary}
For scalar like objects $A$, $B$, $D_i (A) B$ where
\begin{equation}
 D_i (A) := \partial_i - \frac{\partial_i A}{\dot{A}} \frac{\partial}{\partial t},
\end{equation}
is also a scalar like object.

Please notice that $\partial_{\mu} A$, $\partial_{\mu} B$ are vector like objects.
\paragraph{Corollary}
For a scalar like object $A$, 
\begin{equation}
 \partial_i A + \frac{\dot{A}}{\alpha^2} (v_i + \beta_i)
\end{equation}
is also a scalar like object.

Please notice that the above quantity can be written as
$\partial_i A - (\dot{A} / u_0) u_i$ where $u_{\mu}$ is the velocity vector
of the total system and that $\partial_{\mu} A$ is a vector like object.
\paragraph{Proposition $3$}
In the background level, all the evolution equations and all the constraints
can be expressed in the form that polynomials of the scalar like objects only
are vanishing.

As the proof, we write down the Einstein equations.  
As for the space-space components of the metric tensor, we use the matrix notation:
$M := (\tilde{\gamma}_{ij})$, $M^{-1} := (\tilde{\gamma}^{ij})$.
$H$ is the Hubble parameter defined by $\dot{a}/a$
The Einstein equations $G_{\mu \nu} = \kappa^2 T_{\mu \nu}$ give the Hamiltonian 
constraint
\begin{equation}
 \left( \frac{1}{\alpha} H \right)^2 = \frac{\kappa^2}{3} \rho
 + \frac{1}{24} {\rm tr} 
 \left( \frac{1}{\alpha} \dot{M} M^{-1} \frac{1}{\alpha} \dot{M} M^{-1}
 \right),
\label{Hconstraint}
\end{equation}
and the evolution equations
\begin{equation}
 \frac{1}{\alpha} \frac{\partial}{\partial t}
 \left( \frac{H}{\alpha} \right) =
 - \frac{1}{8} {\rm tr} 
 \left( \frac{1}{\alpha} \dot{M} M^{-1} \frac{1}{\alpha} \dot{M} M^{-1}
 \right)
 - \frac{\kappa^2}{2} (\rho + P),
 \label{evolution1}
\end{equation}
\begin{equation}
  \frac{1}{\alpha} \frac{\partial}{\partial t}
 \left( \frac{1}{\alpha} \frac{\partial M}{\partial t}\right) 
 + 3 \frac{H}{\alpha} \left( \frac{1}{\alpha} \dot{M} \right)
 - \frac{1}{\alpha} \dot{M} M^{-1} \frac{1}{\alpha} \dot{M}
 = 0,
 \label{evolution2}
\end{equation}
and the momentum constraint
\begin{eqnarray}
 0 &=& \frac{1}{2} \frac{\dot{a}}{\alpha} 
 D_i (a) \left(\frac{\alpha}{\dot{a}}\right)
 \left( M^{-1} \frac{1}{\alpha} \dot{M} \right)^i_{\; j}
 + \frac{1}{2} 
 \left( M^{-1} D_i (a) M \cdot M^{-1} 
 \frac{1}{\alpha} \dot{M} \right)^i_{\; j} \nonumber\\
 && - \frac{1}{2} 
 \left[ M^{-1} \frac{1}{\alpha} \partial_t 
 \left\{ D_i (a) M  \right\} \right]^i_{\; j}  
 + \frac{1}{4} {\rm tr} 
 \left( M^{-1} D_j (a) M \cdot M^{-1} \frac{1}{\alpha} \dot{M}
 \right) \nonumber\\
  && + 2 D_j (a) \left( \frac{H}{\alpha} \right)
  - \kappa^2 h \frac{\alpha}{a H} Z_j,
 \label{Mconstraints}
\end{eqnarray}
where $Z_i$ is the scalar like object defined by
\begin{equation}
 Z_i := \partial_i a + \frac{\dot{a}}{\alpha^2}
 (v_i + \beta_i). 
\end{equation}
$\Box \phi_a - \partial U / \partial \phi_a = S_a$ gives
\begin{equation}
 \frac{1}{\alpha} \frac{\partial}{\partial t}
 \left( \frac{1}{\alpha} \frac{\partial \phi_a }{\partial t}\right) 
 + 3 \frac{H}{\alpha} \frac{\dot{\phi}_a}{\alpha} 
 + \frac{\partial U}{\partial \phi_a} + S_a
 = 0.
 \label{scalarevolution}
\end{equation}
As for the perfect fluid components, 
$\nabla_{\mu} T^{\mu}_{\alpha 0} = Q_{\alpha 0}$ and
$\nabla_{\mu} T^{\mu}_{\alpha i} = Q_{\alpha i}$ give
\begin{equation}
 \frac{1}{\alpha} \dot{\rho}_{\alpha} = - 3 \frac{H}{\alpha} 
 (\rho_{\alpha} + P_{\alpha}) + Q_{\alpha},
 \label{fluidenergy}
\end{equation}
and
\begin{eqnarray}
 0 &=& \frac{1}{\alpha a^3} 
 \left[ a^2 h_{\alpha} \frac{\alpha}{H} Z_{\alpha i}
 \right]^{\cdot} + D_i (a) P_{\alpha}
 + h_{\alpha} a \frac{H}{\alpha} 
 D_i (a) \left( \frac{\alpha}{\dot{a} }\right)
\nonumber\\
&& - \frac{1}{a} \frac{\alpha}{H} Q_{\alpha} Z_i
- f_{\alpha i}
\label{fluidmomentum} 
\end{eqnarray}
where $h_{\alpha} :=\rho_{\alpha} +P_{\alpha}$ is the enthalpy of the 
fluid component $\alpha$ and $Z_{\alpha i}$ is the scalar like object 
defined by
\begin{equation}
 Z_{\alpha i} := \partial_i a + \frac{\dot{a}}{\alpha^2}
 (v_{\alpha i} + \beta_i). 
\end{equation}
Then we conclude that all the evolution equations and all the constraints
can be written in terms of the scalar like objects only.
The evolution equations of $M$ (\ref{evolution2}) can be solved as
\begin{equation}
 M = R_1 \exp{ \left[ \int_{t_0} dt \frac{\alpha}{a^3} R_2 \right] },
 \label{solutionofM}
\end{equation}
where $R_1$, $R_2$ are the $3 \times 3$ time independent matrices depending on
$\bm{x}$: $R_1$ is unimodular symmetric, $R_2$ is traceless and $R_1 R_2$ 
is symmetric. \cite{Hamazaki2008.2}  
By using(\ref{solutionofM}), the term in (\ref{Hconstraint})(\ref{evolution1})
can be written as   
\begin{equation}
 \frac{1}{4} {\rm tr} 
 \left( \frac{1}{\alpha} \dot{M} M^{-1} \frac{1}{\alpha} \dot{M} M^{-1}
 \right) = \frac{c_R}{a^6}
\label{cRtrace} 
\end{equation}
where
\begin{equation}
 c_R := \frac{1}{4} {\rm tr} (R^2_2). 
\end{equation}

We consider the perturbation.
We assume that the arbitrary background quantity $A$ depends not only
on ($t$, $\bm{x}$), but also on $\lambda$ which characterizes the perturbation.
We can Taylor expand $A$ around $\lambda = 0$ as 
\begin{equation}
 A (\lambda = 1 ) = \sum^{\infty}_{k=0} \frac{1}{k!}
\left. \frac{d^k A(\lambda)}{d \lambda^k} \right|_{\lambda = 0}, 
\end{equation}
where $A (\lambda = 1 )$ is a full nonlinear quantity.
We can identify
\begin{equation}
 \left. \frac{d^k A(\lambda)}{d \lambda^k} \right|_{\lambda = 0}
 \leftrightarrow \delta^k A,
\end{equation}
where $\delta^k A$ is the $k$-th order perturbation of $A$.
The gauge transformation of the background quantity $A$ is defined by
\begin{equation}
 A (\lambda, \mu) = \exp{[\mu L(T)]} A (\lambda, \mu=0), 
 \label{gaugetransso}
\end{equation}
where $L(T)$ is the Lie derivative generated by 
the infinitesimal displacement $T:= T^{\mu} \partial_{\mu}$,
$A (\lambda, \mu=0)$ is the quantity before the gauge transformation
and $A (\lambda, \mu=1)$ is the quantity after the gauge transformation.
This expression (\ref{gaugetransso}) is a solution of the differential 
equation 
\begin{equation}
 \frac{d}{d \mu} A = L(T) A,
\label{gaugetranseq} 
\end{equation}
which we use instead of the solution (\ref{gaugetransso}) from now on.
By differentiating (\ref{gaugetranseq}) with respect to $\lambda$, we get
\begin{equation}
 \frac{d}{d \mu} \frac{d A}{d \lambda} = 
 L \left( \frac{d T}{d \lambda} \right) A
 + L(T) \frac{d A}{d \lambda},
\end{equation}
since not only the background quantity $A$ but also the infinitesimal displacement 
generating the Lie derivative $T$ depends upon $\lambda$.
In general, the gauge transformation of the $\lambda$ derivative of $A$
contain the Lie derivatives generated by $d^k T / d \lambda^k$.  
But we can make a new quantity $B$ by combining the $\lambda$ derivatives
of the background quantity appropriately, so that
\begin{equation}
 \frac{d}{d \mu} B = L(T) B
\label{backgroundlikeobject} 
\end{equation}
which does not contain the Lie derivatives generated by $d^k T / d \lambda^k$
($k=1,2,\cdots$) can hold.
So we can put the definition as follows.
\paragraph{Definition}
We call a quantity $B$ which has the gauge transformation 
(\ref{backgroundlikeobject}) the background like object.

Any background like object is a gauge invariant quantity with respect to
all the infinitesimal gauge transformation satisfying $T(\lambda = 0)=0$.
We can prove the following proposition.
\paragraph{Proposition $4$}
Let $A$, $B$ be the scalar like objects and the background like objects.
Then $D(A) B$ where
\begin{equation}
 D(A) := \frac{d}{d \lambda} - \frac{d A}{d \lambda} \frac{1}{\dot{A}}
 \frac{d}{d t},
\end{equation}
is also a scalar like object and a background like object.

\paragraph{Corollary}
Under the assumptions in the previous proposition, $D(A)^n B$ ($n=1,2,\cdots$)
are also scalar like objects and background like objects.

For the proofs, please see our previous paper. \cite{Hamazaki2008.2}

\paragraph{Proposition $5$}
All the perturbation equations of the evolution equations and the constraints 
can be expressed in the form that the polynomials of the quantities which
are the scalar like objects and the background like objects are vanishing.
That is, all the perturbation equations of the evolution equations and 
the constraints can be expressed in a manifestly gauge invariant manner.

The perturbation equations can be obtained by operating $D(S)$ where $S$ is 
an arbitrary scalar like object on
 (\ref{Hconstraint}) (\ref{evolution1}) (\ref{evolution2}) 
 (\ref{Mconstraints}) (\ref{scalarevolution}) (\ref{fluidenergy})
 (\ref{fluidmomentum})
 which are written in terms of the scalar like objects.
 Since a scalar like object operated $D(S)$ on is a scalar like object and 
a background like object, the assertion of the proposition $5$ follows.
If we want to move the time derivative to the outermost position,
you can use 
\begin{equation}
 \left[ D(S), \frac{1}{\alpha} \frac{\partial}{\partial t}
 \right] = - \frac{\dot{S}}{\alpha} D(S) 
 \left( \frac{\alpha}{\dot{S}} 
 \right) \frac{1}{\alpha} \frac{\partial}{\partial t},
\end{equation}
where $[A,B]:= AB-BA$.

\subsection{the junction conditions}

In the early universe, there exist periods when the equation of state changes quite
rapidly, such as the slow rolling-oscillatory transition and the reheating transition
\cite{Hamazaki1996}.
As the zeroth order approximation, it is appropriate to treat these transitions 
by connecting two spacetimes which have different equations of state by the metric 
junction formalism \cite{Israel1966}. 
These transition hypersurfaces are defined by the particular equations;
for the slow rolling-oscillatory transition, $H/\alpha =m$, and
for the reheating transition $H/\alpha =\Gamma$ where $m$, $\Gamma$ are the mass,
the decay constant of the scalar field, respectively.
Motivated by the above point, we extend the metric junction theory across the 
spacelike hypersurface defined by $C=0$ where $C$ is the scalar like object,
within the framework of the full nonlinear perturbation theory in the leading
order of the gradient expansion.
In the appendix $B$, we formulate the metric junction in the linear perturbation
theory in the long wavelength limit, and mention its consistency with the full
nonlinear theory.

We consider a $4$ dimensional spacetime ${\cal M}$ and a $3$ dimensional hypersurface
$\Sigma$.
$\Sigma$ separates ${\cal M}$ into two region: ${\cal M}_+$ which is the future of
$\Sigma$ and ${\cal M}_-$ which is the past of $\Sigma$.
The hypersurface $\Sigma$ is parametrized by the intrinsic coordinate $y^i$ 
($i=1,2,3$) as
\begin{eqnarray}
 x^0_{\pm} &=& t_{\times} + \delta Z_{\pm} (y),\\
 x^i_{\pm} &=& y^i + \delta Z^i_{\pm} (y),
\end{eqnarray}
where $x^{\mu}_{\pm}$ are spacetime coordinates in the region ${\cal M}_{\pm}$,
respectively and $t_{\times}$ is a constant common to ${\cal M}_{\pm}$.
From now on, we omit index $\pm$.
The gauge transformation of $\delta Z$, $\delta Z^i$ are given by
\begin{eqnarray}
 L(T) \delta Z &=& - T^0,\\
 L(T) \delta Z^i &=& -T^i.
\end{eqnarray}
\paragraph{Proposition $6$}
Let $A_{\mu}$, $B_{\mu}$ be vector like objects.
Then $A_0 \partial_i \delta Z + (A_0 / B_0) B_i$ is the scalar like object.

\paragraph{Corollary}
$\phi_i := \alpha \partial_i \delta Z + (\alpha / \dot{a}) \partial_i a$
is a scalar like object.

In the previous proposition,
as $A_{\mu}$, $B_{\mu}$, please adopt $u_{\mu}$, $\partial_{\mu} a$,
respectively.

As the junction hypersurface, we adopt the hypersurface characterized by
$C=0$ where $C$ is a scalar like object.
Then we get 
\begin{equation}
 \partial_i \delta Z = - \frac{\partial_i C}{\dot{C}},
\end{equation}
which yields
\begin{equation}
 \phi_i = - \frac{\alpha}{\dot{C}} D_i (a) C.
\end{equation}
The normal vector $n_{\mu}$ of the hypersurface $\Sigma$ pointing
from ${\cal M}_-$ to ${\cal M}_+$ is given by
\begin{equation}
 n_{\mu} = 
 \frac{- {\rm sgn}(\dot{C})}
 {\sqrt{ - g^{\rho \sigma} \partial_{\rho} C \partial_{\sigma} C} }
 \partial_{\mu} C,
\end{equation}
and the tangential vector $e^{\mu}_i$ on $\Sigma$ are given by
\begin{equation}
 e^{\mu}_i = \frac{\partial x^{\mu}}{\partial y^i} \quad \quad \quad
 (i=1,2,3).
\end{equation}
Then we get
\begin{equation}
 n_{\mu} n^{\mu} = -1, \quad \quad n_{\mu} e^{\mu}_i=0.
\end{equation}
We define the intrinsic metric $q_{ij}$ and the extrinsic curvature
$K_{ij}$ of $\Sigma$ by
\begin{eqnarray}
 q_{ij} &:=& e^{\mu}_i e^{\nu}_j (g_{\mu \nu} + n_{\mu} n_{\nu}),\\
 K_{ij} &:=& e^{\mu}_i e^{\nu}_j \nabla_{\mu} n_{\nu}.
\end{eqnarray}
In case of $\Sigma$ defined by $C=0$, we obtain
\begin{eqnarray}
 q_{ij} &=& \gamma_{ij} + O(\epsilon^2),\\
 K_{ij} &=& \frac{1}{2 \alpha} \dot{\gamma}_{ij} + O(\epsilon^2).
\end{eqnarray}
As for the energy momentum tensor $T_{\mu \nu}$, we obtain
\begin{eqnarray}
 T_{nn} &:=& n^{\mu} n^{\nu} T_{\mu \nu} = \rho + O(\epsilon^2),\\
 T_{ni} &:=& n^{\mu} e^{\nu}_i T_{\mu \nu} = - (\rho + P) 
  \frac{\alpha}{\dot{a}} \{Z_i - D_i (C)a \} +O(\epsilon^3),\\
 T_{ij} &:=& e^{\mu}_i e^{\nu}_j T_{\mu \nu} = P \gamma_{ij} 
 +O(\epsilon^2).
\end{eqnarray}
We notice that $q_{ij}$, $K_{ij}$, $T_{nn}$, $T_{ni}$ and $T_{ij}$
can be written by the scalar like objects only.
The junction condition formulated by Israel \cite{Israel1966}
is given by
\begin{equation}
 [q_{ij}]^+_- = [K_{ij}]^+_- = [T_{nn}]^+_- = [T_{ni}]^+_- = 0, 
\end{equation}
where $[Q]^+_- := Q_+ - Q_-$.
In our notation, the above juction condition is written by
\begin{equation}
 [a]^+_- =[\tilde{\gamma}_{ij}]^+_- =\left[ \frac{\dot{a}}{\alpha} \right]^+_-
 = \left[ \frac{\dot{\tilde{\gamma}}_{ij}}{\alpha} \right]^+_-
 = [\rho]^+_- = [(\rho + P) \{Z_i - D_i (C)a \} ]^+_- =0.
 \label{junctioncondition}
\end{equation}
\paragraph{Proposition $7$}
In the background level, the metric junction condition can be expressed in  
terms of the scalar like objects only.

We consider the perturbation of the junction condition.
As for the perturbation, the next proposition is essential.
\paragraph{Proposition $8$}
Let the matching hypersurface be defined by $C=0$ where $C$ is a scalar like object.
For an arbitrary scalar like object $S$ satisfying $[S]^+_- =0$,
$D(C) S$, $D_i(C) S$ are continuous across the matching hypersurface:
$[ D(C) S ]^+_- =0$, $ [ D_i(C) S ]^+_-=0$.

For the proof, please see the appendix $A$. 
By applying the above proposition finite times, we obtain the following 
corollary.

\paragraph{Corollary}
Under the assumption presented by the previous proposition,
$[ D(C)^n S ]^+_- =0$ for an arbitrary natural number $n$.

As for $M:=(\tilde{\gamma}_{ij})$, $M$ is solved as 
(\ref{solutionofM}).
From (\ref{junctioncondition}), $M$ in ${\cal M}_+$ is given by
\begin{equation}
 M_+ = R_{1-} \exp{\left[ 
 \int^{t_{\times} + \delta Z_{-}}_{t_0} dt \frac{\alpha}{a^3} R_2
+ \int^t_{t_{\times} + \delta Z_{+}} dt \frac{\alpha}{a^3} R_2
\right]},
\label{nonlinearadiabaticdecaying}
\end{equation}
where $R_{1-}$ is $R_1$ in ${\cal M}_-$, and $R_2$ in ${\cal M}_+$ and
$R_2$ in ${\cal M}_-$ is the same $R_{2 +}=R_{2 -} =: R_2$.

 As the junction, we consider the transition where the energy $\rho_{A-}$ 
transfers into the energy $\rho_{A+}$ which has the different equation 
of state from $\rho_{A-}$. 
From (\ref{junctioncondition}), the energy momentum conservation
\begin{eqnarray}
 &&[\rho_A ]^+_- =0,\\
 &&[(\rho_A + P_A) \{Z_{Ai} - D_i (C)a \} ]^+_- =0, 
\end{eqnarray}
must hold.
In the above discussion, all the perturbation equations of the metric junction
conditions are written in the form that the polynomials of the quantities which 
are the scalar like objects and the background like objects are vanishing,
therefore all the perturbation equations of the metric junction conditions 
are gauge invariant.

\section{choice of the independent gauge invariant variables based on the classification
of the perturbation solutions into the adiabatic mode and the entropic modes}

\subsection{the universal adiabatic growing mode}

All the evolution equations of the locally homogeneous universe are invariant under
the transformation defined by
\begin{eqnarray}
 a &\to& a \Lambda,\\
 R_2 &\to& R_2 \Lambda^3,\\
 \alpha &\to& \alpha,\\
 W &\to& W, 
\end{eqnarray}
where $\alpha$ is the lapse function and $W$ is an arbitrary scalar quantity.
$\Lambda$ is a time independent function $\Lambda = \Lambda ({\bm x})$.
Taking the variation with respect to $\lambda$ considering that only $\Lambda$
is dependent upon $\lambda$ gives the obvious perturbation solution.
We call this solution the universal adiabatic growing mode. \cite{Kodama1998} 
For an arbitrary scalar quantity $S$, the first order and the second order
perturbation solutions of the universal adiabatic growing mode is written as
\begin{eqnarray}
 D(a) S &=& - \frac{\dot{S}}{H} \frac{d}{d \lambda} \ln{\Lambda},\\ 
 D(a)^2 S &=& - \frac{\dot{S}}{H} \frac{d^2}{d \lambda^2} \ln{\Lambda}
 + \frac{1}{H} \frac{d}{d t} \left( \frac{\dot{S}}{H} \right) 
 \left( \frac{d}{d \lambda} \ln{\Lambda} \right)^2,
\end{eqnarray}
where the first order expression is very familiar in large literature.
We call the gauge invariant perturbation variable defined by
\begin{equation}
 \zeta_n (S) := D(S)^n \ln{a}
\end{equation}
the generalized Bardeen parameter induced by the scalar like object $S$.
When we adopt an arbitrary scalar quantity $W$ or $H / \alpha$ as the scalar
like object $S$, the perturbation solution of the universal adiabatic growing
mode is written as a time independent form:
\begin{equation}
 \zeta_n (S) =
 \frac{d^n}{d \lambda^n} \ln{\Lambda}. 
\end{equation}

\subsection{the adiabatic perturbation variable and the entropic perturbation variables}

In order to interpret the physics of the linear cosmological perturbations, the classification
into the adiabatic perturbation and the entropic perturbations was often convenient.
\cite{Kodama1984} \cite{Kodama1987} \cite{Mukhanov1992} \cite{Gordon2001}
Therefore the generalization of this classification into higher order perturbations are 
thought to be useful.
So we define the adiabatic perturbation variable and the entropic perturbation variables
in the higher order perturbation theory.
\paragraph{Definition}
We call the perturbation variable which does not vanish for the universal adiabatic growing
mode the adiabatic perturbation variable. 
We call the perturbation variable which vanishes for the universal adiabatic growing
mode the entropic perturbation variable.

We will present the examples of the adiabatic and the entropic perturbation variables.
We assume that $S$, $S_i$ ($i=1,2$) are the scalar like objects such as $W$, 
$\dot{W}/\alpha$, $H / \alpha$ where $W$ is an arbitrary scalar variable.
The generalized Bardeen parameter $\zeta_n (S)$ and $D(a)^n S$ are adiabatic perturbation 
variables and $\zeta_n (S_1)-\zeta_n (S_2)$ and $D(S_1)^n S_2$ are entropic perturbation variables.

\subsection{the $N$ philosophy and the $\rho$ philosophy}

We call the expressions representing the physical quantities at the final time in terms
of those at the initial time the $S$ formulas. \cite{Hamazaki2008}
Our final purpose is to construct the $S$ formulas of the adiabatic perturbation 
variable such as the Bardeen parameter $\zeta_n (\rho) := D(\rho)^n \ln{a}$.
In the previous subsection, it was shown that the Bardeen parameter $\zeta_n (\rho)$
is time independent for the universal adiabatic growing mode.
Therefore we expect that the formulation in which the difference between the Bardeen 
parameter at the final time and that at the initial time can be expressed in terms
of the entropic perturbation variables may exist.
In this subsection, we choose the appropriate set of the entropic perturbation variables
and we construct the formulation in which the time change of the Bardeen parameter 
is brought about by the evolutions of the set of these entropic perturbation variables.

Until now in order to understand the evolutions of linear cosmological perturbations 
in the universe governed by the multiple component energy densities, the decomposition
of the perturbations into the adiabatic component and the entropic components has
already been performed.\cite{Kodama1984} \cite{Kodama1987} \cite{Mukhanov1992} \cite{Gordon2001}
In the nonlinear perturbation theory, the following set of perturbation variables was 
adopted: as the adiabatic perturbation variable, the Bardeen parameter $\zeta_n (\rho)$,
and as the entropic perturbation variables, the difference between the generalized 
Bardeen parameters induced by the energy densities of the different components
$S_n (\rho_A, \rho_B) := \zeta_n (\rho_A) - \zeta_n (\rho_B)$ where the subscripts $A$, $B$
represent the different components.\cite{Langlois2008}
Since all the perturbation variables in this formulation \cite{Langlois2008} are
based on perturbations of the logarithm of the scale factor $N:=\ln{a}$, we call
this formulation the $N$ philosophy.
But in the $N$ philosophy, it is difficult to write down the evolution equations in terms
of the set of variables $\zeta_n (\rho)$, $S_n (\rho_A, \rho_B)$ in the closed form.
From now on, we often consider the matching of the metric across the matching hypersurface
defined by $\rho = {\rm const}$.
Across such matching hypersurface,
the perturbation variables in the $N$ philosophy, $S_n (\rho_A, \rho_B)$ jump by finite values.

In order to solve the defects in the $N$ philosophy, we propose the new set of the 
perturbation variables.
We choose
\begin{equation}
 D \left( \frac{H}{\alpha} \right)^n \ln{a}, \quad \quad
 D \left( \frac{H}{\alpha} \right)^n s_A,
\label{originalrhophilosophy} 
\end{equation}
where $s_A := \rho_A / \rho$, as the adiabatic perturbation variable and the entropic
perturbation variables, respectively.
$s_A$ satisfies $\sum_A s_A = 1$.
As for the new set of the perturbation variables, no finite jumps do not exist across
the slow rolling-oscillatory transition $H/ \alpha =m$ where $m$ is the mass of the 
scalar field and across the reheating transition $H/ \alpha =\Gamma$ where $\Gamma$
is the decay constant of the scalar field.
In order to avoid the calculational complexity, we assume that $c_R =0$, since
we can neglect the second term of the right hand side of the Hamiltonian constraint 
(\ref{Hconstraint}) with (\ref{cRtrace}) because of the rapid growth of the 
scale factor $a$ during the inflationary expansion of the universe.
In the condition $c_R =0$, the matching conditions of the slow rolling-oscillatory
transition, of the reheating transition are reduced into  
$\rho - 3 m^2 / \kappa^2 =0$, $\rho - 3 \Gamma^2/ \kappa^2 =0$, respectively.
Under the simplification of $c_R =0$, the set of the perturbation variables which 
we adopted in (\ref{originalrhophilosophy}) is reduced into 
\begin{equation}
 D ( \rho )^n \ln{a}, \quad \quad
 D ( \rho )^n s_A.
\label{rhophilosophy} 
\end{equation}
Since these perturbation variables are continuous across the matching hypersurface 
defined by $\rho={\rm const}$, we only have to concentrate on solving the 
evolution equations of these perturbation variables.
Since all these variables are defined by $D(\rho)$, we call the use of these 
variables presented in (\ref{rhophilosophy}) the $\rho$ philosophy.

We will give the evolution equations of the perturbation variables 
(\ref{rhophilosophy}).    
For simplicity, we assume that the multiple components do not interact and 
that $\rho_A$ obeys $d \rho_A / d N = - g_A \rho_A$ where $g_A$ will be called
the $g$ factor from now on.
When the $\alpha$ component is the perfect fluid with 
$w_{\alpha} := P_{\alpha} / \rho_{\alpha}$,
its $g$ factor is given by $g_{\alpha} = 3 (1+w_{\alpha})$.  
When the $a$ component is the slow rolling massive scalar field with mass $m_a$,
its $g$ factor is given by $g_a = 2 m^2_a / \kappa^2 \rho$.
Since it was shown that the oscillatory massive scalar field can be approximated
by the perfect fluid with $w_{\alpha} =0$ \cite{Kodama1996} \cite{Hamazaki2002}
\cite{Hamazaki2004} \cite{Hamazaki2008}, we can use the perfect fluid with 
$g_{\alpha}=3$ instead of the oscillatory massive scalar field.
The evolution equations of $N := \ln{a}$, $s_A$ are given by
\begin{eqnarray}
 \frac{d}{d \rho} N &=& - \frac{1}{\rho s},
 \label{evolutionrhoN}\\
 \frac{d}{d \rho} s_A &=& \frac{1}{\rho} \left( - s_A + \frac{g_A s_A}{s} \right),
 \label{evolutionrhosa} 
\end{eqnarray}
where $s := \sum_B g_B s_B$.
We choose the total energy density $\rho$ as the evolution parameter instead of the 
cosmic time $t$, and the right hand sides of (\ref{evolutionrhoN}), (\ref{evolutionrhosa})
are written by $\rho$, $s_A$ only. 
The evolution equations of the perturbation variables in the $\rho$ philosophy
are given by operating $D(\rho)$ finite times on (\ref{evolutionrhoN}), 
(\ref{evolutionrhosa}).
In this case, it is important to notice that $D(\rho)$ and $d / d \rho$ are commutative 
since $d / dt$ and $d / d \lambda$ are commutative.

In the $\rho$ philosophy, all the perturbation variables are continuous across the 
matching hypersurface defined by $\rho={\rm const}$ because of the proposition $8$, 
and we can easily derive the evolution equations of the perturbation variables.
Since the $\rho$ philosophy is superior to the $N$ philosophy because of the above two
reasons, we will adopt the $\rho$ philosophy from now on.

\section{the non-Gaussianities of the nonlinear cosmological perturbations}

In this section, we discuss the non-Gaussianities generated in several cosmological
models.
The non-Gaussianities are measured by the $f_{NL}$ parameter.\cite{Komatsu2001}
It is assumed that the logarithm of the scale factor $N := \ln{a}$ is given by
the function of the energy density $\rho$ as the evolution parameter and of the 
solution constants.
We only consider the models where the origins of the cosmological perturbations 
are in the quantum fluctuations of the scalar fields $\phi_a$ in the inflationary
universe.
The statistical mean values of the perturbation amplitudes are given by
\begin{equation}
 \left<\left< \frac{d \phi_a(0)}{d \lambda} \frac{d \phi_b(0)}{d \lambda} 
 \right>\right> \sim H^2 \delta_{ab},
\end{equation}
where $\phi_a(0)$ is the expectation value of the scalar field $\phi_a$ at the 
first horizon crossing and $H$ is the Hubble parameter at the first horizon crossing.
In this case, the solution constants are given by the set of the expectation values
of the scalar fields at the first horizon crossing $\{ \phi_a (0) \}$.
In this case, using the logarithm of the scale factor 
$N=N(\rho, \phi_1(0), \phi_2 (0), \cdots)$ 
the non-Gaussianity parameter $f_{NL}$ is defined by
\begin{equation}
 f_{NL} := \frac{N_{ab} N^a N^b}{(N_c N^c)^2},
\end{equation}
\begin{equation}
 N_a := \frac{\partial}{\partial \phi_a (0)} N, \quad \quad
 N_{ab} := \frac{\partial^2}{\partial \phi_a (0) \partial \phi_b (0) } N.
\end{equation}
\cite{Komatsu2001}
$N_a$, $N_{ab}$ are defined as the coefficients given when we expand
the gauge invariant adiabatic perturbation variables $D(\rho) \ln{a}$,
$D(\rho)^2 \ln{a}$ with respect to $d \phi_a (0) / d \lambda$, respectively;
\begin{align}
 D(\rho) \ln{a} &= \sum_a N_a \frac{d \phi_a (0)}{d \lambda},\\
 D(\rho)^2 \ln{a} &= \sum_{ab} N_{ab} \frac{d \phi_a (0)}{d \lambda}
 \frac{d \phi_b (0)}{d \lambda}.
\end{align}
We assumed that the more than second order perturbations of $\phi_a$ at 
the initial time are all vanishing;
$d^n \phi_a (0) / d \lambda^n =0$ ($n \ge 2$).

Since the cosmological perturbations have the origin in the scalar field fluctuations 
in the de Sitter stage, the deviations of the spectral indices of the Bardeen parameter
$\zeta_1 (\rho)$ from the scale invariance $d \ln{<< \zeta_1 (\rho) >>} / d \ln{k}$ are 
suppressed by the slow rolling parameter.
Since the first horizon crossing is defined by the relation
\begin{equation}
 k = a H = e^{N_{\ast}} \frac{\kappa}{\sqrt{3}} \rho(0)^{1/2},
\end{equation}
where $N_{\ast}$, $\rho(0)$ is the logarithm of the scale factor, energy density
at the first horizon crossing time, respectively, we obtain
\begin{equation}
 d \ln{k} = \left\{ 1+ \frac{1}{2} \frac{1}{\rho(0)} \frac{d \rho(0)}{d N_{\ast}}
 \right\} d N_{\ast}.
\end{equation}
The slow rolling phase is characterized by the smallness of the $g$ factors of the 
scalar fields $\phi_a (0)$ whose size is bounded by $\delta_S$ a small constant
characterizing the slow rolling of the scalar fields;
$|g_a| \le \delta_S$ where $g_a$ is defined by the evolution equations of the 
energy densities of the scalar field $\phi_a$; $\rho_a$ at the first horizon 
crossing: $d \rho_a (0) /d N_{\ast}= - g_a \rho_a(0)$.
In all the cases which we consider, the following evaluations hold:
$\partial \ln{<<\zeta^2_1 (\rho)>>} / \partial \rho_a (0) \sim 1/ \rho_a (0)$.
By using the above properties, we can conclude that the Bardeen parameter in the 
first order perturbation theory $\zeta_1 (\rho)$ has almost complete scale 
invariance:
\begin{equation}
 \frac{d}{d \ln{k}} \ln{<< \zeta^2_1 (\rho) >>} \sim \delta_S.
\end{equation}
Then we concentrate on the non-Gaussianity parameter $f_{NL}$ from now on.

Except for the cases where the large $f_{NL}$ is generated discontinuously
on the transition hypersurface such as the inhomogeneous end of the inflation
\cite{Lyth2005.2} \cite{Sasaki2008} and the modulated reheating \cite{Sasaki1991}
\cite{Dvali2004}, different two cases have been discussed.
One is the curvaton scenario \cite{Ichikawa2008} and the other is the vacuum 
dominated two scalar fields \cite{Alabidi2006} \cite{Byrnes2009}.
We discuss that the mechanism which generates the large $f_{NL}$ continuously
in the above two different cases can be explained from the three common viewpoints
which will be presented in the subsection $2$ of this section.
In this section, we use two strong methods; the exponent evaluation method
and the $\rho$ philosophy.
By the exponent evaluation method, it becomes possible to evaluate the order
of $f_{NL}$ analytically in the wide ranges of parameters.
In the $\rho$ philosophy, the time evolutions of the above two systems are traced
using the logarithm of the scale factor $N$ as the adiabatic independent variable,
$s_2$ which implies the ratio of the energy density of the component which does not
govern the energy density of the universe as the entropic independent variable, 
and the energy density $\rho$ as the evolution parameter.
The $\rho$ philosophy makes the instant when $f_{NL}$ grows large and the length
of the period when the large $f_{NL}$ continues clear.

\subsection{the exponent evaluation method}

In many papers, the calculations of the non-Gaussianity parameter $f_{NL}$ were
performed: for the inhomogeneous end of the inflation \cite{Lyth2005.2}
\cite{Sasaki2008}, for the modulated reheating \cite{Dvali2004}, for the 
curvaton \cite{Ichikawa2008} and for the vacuum dominated inflation represented
by the hybrid inflation \cite{Alabidi2006} \cite{Byrnes2009}.  
In many papers so far, the calculations of $f_{NL}$ were performed numerically
and in the extreme situations where only one factor is concerned the analytic 
formulas of $f_{NL}$ were given.
In this subsection, we present the exponent evaluation method by which it 
becomes possible to give the analytic order estimates of $f_{NL}$ in the 
wide range including cases where more than two factors are concerned.

We explain the exponent evaluation method by adopting the inhomogeneous end of the 
inflation \cite{Lyth2005.2} \cite{Sasaki2008} as an example.
We consider the two scalar fields $\phi_1$, $\phi_2$ governed by the 
vacuum dominated potential given by
\begin{equation}
 U = U_0 + \sum_{a=1}^2 \frac{1}{2} \eta_a \phi_a^2,
\end{equation}
where $\eta_a$ ($a=1,2$) are negative and $U_0$ is the large constant compared 
with the terms quadratic with the scalar fields.
Under the approximation where the vacuum energy $U_0$ is the dominant contribution
of the energy density $\rho$ and where the scalar fields $\phi_a$ ($a=1,2$) are
slow rolling on the potential, the evolutions of $\phi_a$ ($a=1,2$) are given by
\begin{equation}
 \phi_a = \phi_a (0) \exp{\left[ - \frac{\eta_a}{\kappa^2 U_0} N \right]}.
\end{equation}
For simplicity, $\eta_a$ ($a=1,2$) are assumed to be $a$ independent:
$\eta_a = \eta$.
We assume that the inflation ends in the bifurcation set defined by
\begin{equation}
 \sum_a \gamma_a \phi_a^2 = \sigma^2.
\label{bifurcationset} 
\end{equation}
On the bifurcation set, the waterfall field which interacts with the 
inflatons $\phi_a$ ($a=1,2$) gets the negative mass and the large vacuum energy
$\sim U_0$ is transferred into the oscillation energy of the waterfall field 
and into the radiation energy which interacts with the waterfall field.
Also in the situation where along the curve defined by (\ref{bifurcationset})
the deep ditch of the potential exists, the inflation ends 
on the curve (\ref{bifurcationset}).
The logarithm of the scale factor $N$ at which the scalar fields reach the 
bifurcation set (\ref{bifurcationset}) is given by
\begin{equation}
 N = \frac{\kappa^2 U_0}{2 \eta} 
 \ln{\left[ \frac{1}{\sigma^2} \sum_a \gamma_a \phi_a^2 (0) \right]}.
\end{equation}
In order that the inflation can solve the horizon problem, we assume that 
$\alpha := \kappa^2 U_0 / 2 \eta = 10^2$.  
The non-Gaussianity parameter $f_{NL}$ is calculated as  
\begin{equation}
 f_{NL} = \frac{1}{\alpha} \left( \frac{A_1 A_3}{2 A_2^2} -1 \right)
 =: \frac{1}{\alpha} (p-1),
\end{equation}
where
\begin{equation}
 A_n := \sum_{a=1}^2 \gamma_a^n \phi_a^2 (0).
\end{equation}
From now on, we neglect numerical factors of order unity without mentioning.
By setting 
\begin{equation}
 \gamma_r := \frac{\gamma_2}{\gamma_1} = 10^k, \quad \quad
 \phi_r := \frac{\phi_2 (0)}{\phi_1 (0)} = 10^{-l},
\end{equation}
we obtain
\begin{equation}
 p = \frac{(1+10^{k-2l})(1+10^{3k-2l})}{(1+10^{2k-2l})^2}.
\end{equation}
The exponent evaluation method gives
\begin{align}
 &k<\frac{2}{3}l         &p&=1,\\
 &\frac{2}{3}l < k <l    &p&=10^{3k-2l},\\
 &l < k < 2l             &p&=10^{-k+2l},\\
 &2l< k                  &p&=1. 
\end{align}
As an example of application of the exponent evaluation method,
we adopt $l < k <2l$.
Since $l < k <2l$, we obtain 
\begin{align}
 1 + 10^{k-2l} &\sim 1,\\
 1 + 10^{3k-2l}&\sim 10^{3k-2l},\\
 1 + 10^{2k-2l}&\sim 10^{2k-2l},
\end{align}
and get
\begin{equation}
 p= \frac{1 \cdot 10^{3k-2l}}{(10^{2k-2l})^2}=10^{-k+2l}.
\end{equation}
$p$ takes the maximum at $k=l$: $p=10^l$,
then $f_{NL} = 10^{l-2}$.
In the exponent evaluation method, we take only the term which has 
the largest exponent in the polynomial constructed by several $10^M$ type 
terms.

As the second example of application of the exponent evaluation method,
we consider the modulated reheating. \cite{Sasaki1991} \cite{Dvali2004}
The scalar field $\phi_1$ on the potential $U=m^2 \phi_1^2 /2$ causes 
the chaotic inflation:
\begin{equation}
 N = \frac{\kappa^2}{4} \left\{ \phi_1^2 (0) - \phi_1^2 \right\}.
\end{equation}
When $H^2 =m^2$ where $H$ is the Hubble parameter, the scalar field $\phi_1$
begins to oscillate and behaves like a dust fluid \cite{Kodama1996} \cite{Hamazaki2002}
\cite{Hamazaki2004} \cite{Hamazaki2008}:
\begin{equation}
 N = \frac{\kappa^2}{4} \phi_1^2 (0) -\frac{3}{2} 
 -\frac{1}{3} \ln{\left( \frac{\kappa^2 \rho}{3 m^2}\right)}.
\end{equation}
When $H^2 = \Gamma^2$ where $\Gamma$ is the decay constant of the scalar field $\phi_1$,
the scalar field oscillation is transformed into the radiation fluid:
\begin{equation}
 N = \frac{\kappa^2}{4} \phi_1^2 (0) -\frac{3}{2} 
 - \frac{1}{3} \ln{\frac{\Gamma^2}{m^2}}
 -\frac{1}{4} \ln{\left( \frac{\kappa^2 \rho}{3 \Gamma^2}\right)}.
\end{equation}
In the modulated reheating, we consider that the decay constant of the first
scalar field $\phi_1$; $\Gamma$ is the function of the second scalar field $\phi_2$ as
$\Gamma = \alpha_d \phi_2^n (0)$ where $\alpha_d$ is a constant and $n$ is an 
integer.
In this case, $N$ takes the form as
\begin{equation}
 N = \kappa^2 \phi_1^2 (0) + \ln{\phi_2 (0)},
\label{modulatedN}
\end{equation}
up to the $\rho$ dependent part which does not contribute the Bardeen parameters
$\zeta_n (\rho)$.
In order that the $\phi_1$ inflation can solve the horizon problem, we assume 
that $\phi_1 (0) = 10/ \kappa$. 
We assume that the second scalar field $\phi_2 (0)$ takes a small value as
$\phi_2 (0) = 10^{-l}/ \kappa$.
The non-Gaussianity parameter $f_{NL}$ is calculated as
\begin{equation}
 f_{NL} = \frac{10^2 -10^{4l}}{(10^2 + 10^{2l})^2}.
\end{equation}
The exponent evaluation method gives
\begin{align}
 &l < \frac{1}{2}   &f_{NL}&=10^{-2},      \label{modulatedcase1}\\
 &\frac{1}{2} <l <1 &f_{NL}&=-10^{-4(l-1)},\label{modulatedcase2}\\
 &1 < l             &f_{NL}&=-1,           \label{modulatedcase3}
\end{align}
When the second scalar field $\phi_2$ takes a very small value, a significant
non-Gaussianity $f_{NL}$ is generated.

\subsection{the non-Gaussianities induced by the entropic perturbation of the component
which does not govern the cosmic energy density}

In this subsection, we investigate the mechanism which triggers the large $f_{NL}$ in 
the different two models $1$, $2$ where the large $f_{NL}$ can be generated continuously.
The model $1$ is the radiation-dust system and the non-Gaussianity $f_{NL}$ in the 
model $1$ was investigated in the paper \cite{Ichikawa2008} in the context of the 
curvaton scenario.
The model $2$ is the vacuum dominated two scalar fields and the non-Gaussianity $f_{NL}$ 
in the model $2$ was investigated in the papers \cite{Alabidi2006} \cite{Byrnes2009}
in the context of the hybrid inflation.
Although the model $1$ and the model $2$ are quite different apparently, we consider 
the mechanisms which generate the large $f_{NL}$'s in the model $1$ and in the model $2$
are completely the same.
It is assumed that the inflation sufficient to solve the horizon problem $N \sim 10^2$
is brought about by the first scalar field $\phi_1$.
Under this assumption, the common points are summarized in the following three points:

\begin{enumerate}
 \item[A.] The expectation value of the second scalar field $\phi_2$; $\phi_2 (0)$ is very
small compared with that of the first scalar field $\phi_1$; $\phi_1 (0)$ at the first 
horizon crossing. 
 \item[B.] The component which originates from the second scalar field $\phi_2 (0)$ does not
govern the cosmic energy density $\rho$.
 \item[C.] The $g$ factor of the component originating from the second scalar field $\phi_2 (0)$;
$g_2$ is smaller than the $g$ factor of the component governing the cosmic energy density
$\rho$; $g_1$. 
\end{enumerate}
The condition $A$ guarantees that the contribution from the second scalar field $\phi_2 (0)$
to the Bardeen parameters $\zeta_n (\rho)$ ($n=1,2,\cdots$) is large.
For example, we assume that $N$ can be written as the sum of the $\phi_i (0)$ dependent 
parts; $N = \sum_i f_i (\phi_i (0))$ and that each $f_i$ is the power or the logarithm
of $\phi_i (0)$.
Then the contribution from $\phi_2 (0)$ to $\zeta_1 (\rho)$ is proportional to
$\partial N / \partial \phi_2 (0) \sim f_2 (\phi_2 (0))/ \phi_2 (0)$ which is quite large 
when $\phi_2 (0)$ is very small even if the contribution from $\phi_2 (0)$ to $N$; 
$f_2 (\phi_2 (0))$ is quite small.
Owing to the condition $A$, it becomes possible that the entropic perturbation of the component
which does not govern the cosmic energy density $\rho$ brings about significant contributions 
to the Bardeen parameters $\zeta_n (\rho)$ ($n=1,2,\cdots$) without contributing $N$. 
From the structure of the evolution equation of $N:=\ln{a}$ (\ref{evolutionrhoN}) and that of 
$s_A$ (\ref{evolutionrhosa}), the growth of the Bardeen parameters 
$\zeta_n (\rho) := D(\rho)^n N$ are governed by the entropic perturbation 
$D(\rho)^k s_A$ of the component $A$ whose ratio of the energy density $s_A$ is 
very small $|s_A| \ll 1$.
In addition, when $|s_A| \ll 1$, the entropy perturbation $D(\rho)^k s_A$ can grow or 
decrease rather rapidly.
Therefore the condition $B$ requires the existence of such component.
Owing to the condition $C$; $g_2 < g_1$, the contribution of $s_2$ to $N$ is
monotonically increasing for the time evolution when the evolution parameter $\rho$ becomes 
decreasing, therefore the contribution from $D(\rho)^k s_2$ to the Bardeen parameter
$\zeta_n (\rho) := D(\rho)^n N$ is also increasing.

Since $g_2 < g_1$ from the condition $C$, the ratio of the energy density of the second 
component $s_2$ grows compared with that of the first component $s_1$.
Since the second component $s_2$ begins to dominate the cosmic energy density $\rho$ soon,
the period when the condition $B$ is satisfied is only the early period of time from the 
beginning.
So the large $f_{NL}$ is realized only transiently in this period.

Both the model $1$ and the model $2$ share the above three properties.
The model $1$ is treated as the case without the cosmological term in
section $4.2.1$ and the model $2$ is treated as the case with the cosmological term in
section $4.2.2$.

In the rest of this subsection $4.2$, we analyse the concrete physical systems using
$D(\rho)^n \ln{a}$, $D(\rho)^n s_A$ as the independent perturbation variables and 
$\rho$ as the evolution parameter.
This scheme which we proposed as the $\rho$ philosophy in subsection $3.3$ is 
supported by the results of sections $2$, $3$.
Subsection $2.1$ guarantees that these independent perturbation variables
$D(\rho)^n \ln{a}$, $D(\rho)^n s_A$ are gauge invariant perturbation variables.
Subsection $3.2$ states that $D(\rho)^n \ln{a}$, $D(\rho)^n s_A$ can be regarded as
the adiabatic perturbation, the entropic perturbations in the higher order perturbation, 
respectively.
This A/E interpretation is useful when we interpret the time evolutions of the concrete
physical systems.
Since $D(\rho) \rho =0$, that is $D(\rho)$ can be interpreted as the partial derivative 
with respect to $\lambda$ with $\rho$ fixed, the $\rho$ dependences of $\ln{a}$, $s_A$
are directly reflected to the $\rho$ dependences of $D(\rho)^n \ln{a}$, $D(\rho)^n s_A$.
This fact makes the calculations and the interpretations of the time evolutions of 
the perturbation variables transparent.
By the $\rho$ philosophy which we explained in the above, in the rest of this 
subsection $4.2$ we clarify that the entropy perturbation $D(\rho)^n s_2$
supported by the energetically subdominant component $\rho_2$ makes the adiabatic
perturbation $D(\rho)^n \ln{a}$  and the non-Gaussianity $f_{NL}$ grow considerably
under the conditions that the $g$ factor of this energetically subdominant component
$\rho_2$ is smaller than the $g$ factor of the energetically dominant component $\rho_1$
and that the subdominant component $\rho_2$ is supported by the extremely small 
scalar field expectation value.

\subsubsection{the case without the cosmological term}

We consider the two component system.
We assume that the $g$ factor of the component $\rho_A$ ($s_A$) is $g_A$.
Assuming that $|s_2| \ll 1$ and linearizing (\ref{evolutionrhoN})(\ref{evolutionrhosa}) 
with respect to $s_2$, we obtain
\begin{eqnarray}
 \frac{d}{d \rho} s_2 &=& \frac{1}{\rho} \left( -1 + \frac{g_2}{g_1} \right) s_2,
\label{evowithouteq1}\\
 \frac{d}{d \rho} N &=& -\frac{1}{\rho g_1} + \frac{1}{\rho g_1} 
 \left( -1 + \frac{g_2}{g_1} \right) s_2,
\label{evowithouteq2}
\end{eqnarray}
whose solution is given by 
\begin{eqnarray}
 s_2 &=& \frac{\rho_2 (1)}{\rho_1 (1)} 
 \left( \frac{\rho}{\rho_1 (1)} \right)^{-1+g_2/g_1},
 \label{evowithoutsol1}\\
 N &=& N(1) -\frac{1}{g_1} \ln{ \frac{\rho}{\rho_1 (1)} } +
 \frac{1}{g_1} \frac{\rho_2 (1)}{\rho_1 (1)} 
 \left( \frac{\rho}{\rho_1 (1)} \right)^{-1+g_2/g_1},
\label{evowithoutsol2} 
\end{eqnarray}
where the $g$ factors are assumed to be constant and $X(1)$ implies 
the physical quantity $X$ at an initial time.
When $g_2 < g_1$, $s_2$ and the contribution to $N$ from the $\rho_2 (1)$
dependent term increase and they are not bounded for the time evolution 
$\rho \to 0$.
When $g_2 > g_1$, $s_2$ and the contribution to $N$ from the $\rho_2 (1)$
dependent term decrease for the time evolution.

When $g_2 <g_1$, the ratio of the energy density $s_2$ increases and reaches 
almost unity.
In this case, the evolution equations (\ref{evolutionrhoN}) (\ref{evolutionrhosa})
give 
\begin{eqnarray}
 \frac{d}{d \rho} s_1 &=& \frac{1}{\rho} \left( -1 + \frac{g_1}{g_2} \right) s_1,
\label{evowithouteq3}\\
 \frac{d}{d \rho} N &=& -\frac{1}{\rho g_2} + \frac{1}{\rho g_2} 
 \left( -1 + \frac{g_1}{g_2} \right) s_1,
\label{evowithouteq4}
\end{eqnarray}
by linearizing (\ref{evolutionrhoN}) (\ref{evolutionrhosa}) with respect to $s_1$
assuming that $|s_1| \ll 1$.
The solution is given by 
\begin{eqnarray}
 s_1 &=& \frac{\rho_1 (1)}{\rho_2 (1)} 
 \left( \frac{\rho}{\rho_2 (1)} \right)^{-1+g_1/g_2},
 \label{evowithoutsol3}\\
 N &=& N(1) -\frac{1}{g_2} \ln{ \frac{\rho}{\rho_2 (1)} } +
 \frac{1}{g_2} \frac{\rho_1 (1)}{\rho_2 (1)} 
 \left( \frac{\rho}{\rho_2 (1)} \right)^{-1+g_1/g_2}.
\label{evowithoutsol4} 
\end{eqnarray}
When $g_2 < g_1$, $s_1$ and the contribution to $N$ from the $\rho_1 (1)$
dependent term decrease for the time evolution $\rho \to 0$.

In the above, the independent variables $\ln{a}$, $s_A$ are written as 
functions of $\rho$ as the evolution parameter.
By operating $D(\rho)$ derivatives on expressions of $\ln{a}$, $s_A$,
we can obtain the expressions of the A/E perturbation variables
$D(\rho)^n \ln{a}$, $D(\rho)^n s_A$ in the form of the functions of $\rho$.
Please notice that the $D(\rho)$ derivative can be regarded as the partial 
derivative with respect to $\lambda$ with the evolution parameter $\rho$ fixed.
Therefore the $\rho$ dependences of $\ln{a}$, $s_A$ completely
corresponds with the $\rho$ dependences of $D(\rho)^n \ln{a}$, $D(\rho)^n s_A$.
In the above calculations and discussions, we can see that the entropic
perturbation $D(\rho)^n s_2$ of the energetically subdominant component
$\rho_2$ with $g$ factor smaller than the $g$ factor of the energetically 
dominant component $\rho_1$ makes the adiabatic perturbation variable
$D(\rho)^n \ln{a}$ grow transiently while the energy density ratio $s_2$
is increasing.

When we calculate the non-Gaussianity $f_{NL}$, the expressions of $N_a$,
$N_{ab}$ are necessary.
$N_a$, $N_{ab}$ are defined as the coefficients of the gauge invariant
adiabatic perturbation variables $D(\rho) \ln{a}$, $D(\rho)^2 \ln{a}$ 
given when $D(\rho) \ln{a}$, $D(\rho)^2 \ln{a}$ are 
expanded with respect to $d \phi_a (0) / d \lambda$, respectively.
The subscript $a$ implies the $\partial / \partial \phi_a (0)$ derivative
with $\rho$ fixed.
The $\rho$ philosophy proposed in subsection $3.3$ makes the calculations and the
interpretations of the time evolutions of the non-Gaussianity $f_{NL}$
as well as the A/E perturbation variables $D(\rho)^n \ln{a}$, $D(\rho)^n s_A$
more simple and more transparent.

We apply the above results to the following concrete situation.
Before $N=N(1)$, $\rho_1$ causes the chaotic inflation $N(1) \sim 10^2$,
and at $N=N(1)$ decays into the radiation.
After $N=N(1)$, $\rho_1$ is radiation ($g_1 = 4$).
After $N=N(1)$, $\rho_2$ is still in the slow rolling phase.
Because we assume that $m_2 \ll m_1$, $\phi_2$ hardly moves from the initial
value $\phi_2 (0)$ ($g_2 =0$).
At $\kappa^2 \rho / 3 =m^2_2$, the slow rolling phase of $\rho_2$ ends and 
begins to oscillate.
After $\kappa^2 \rho / 3 =m^2_2$, $\rho_2$ behaves like dust fluid ($g_2 =3$).
\cite{Kodama1996} \cite{Hamazaki2002} \cite{Hamazaki2004} \cite{Hamazaki2008}
Then applying (\ref{evowithoutsol2}) to the above situation yields
\begin{equation}
 N = N(1) + \frac{1}{4} \ln{ \frac{\rho_1 (1)}{\rho} } + \frac{1}{8}
 m^2_2 \phi_2^2 (0) \left( \frac{\kappa^2}{3 m_2^2} \right)^{3/4}
 \frac{1}{\rho^{1/4}} \frac{1}{(1 - \kappa^2 \phi_2^2 (0)/6)^{3/4}}.
\end{equation}
This solution is simplified into the model given by
\begin{equation}
 N = \kappa^2 \phi_1^2 (0) + \kappa^2 \alpha(\rho) \phi_2^2(0),
\label{Nmodel1} 
\end{equation}
up to the $\rho$ dependent part which is not related with the Bardeen parameters
$\zeta_n (\rho)$, where $\alpha(\rho)$ is a function of $\rho$ and increases for 
the time evolution $\rho \to 0$.
All the numerical coefficients of order unity are dropped without mentioning from
now on.
We assume that $\phi_1$ causes the inflation enough to solve the horizon problem
and that $\phi_2$ is a small field enough to contribute to the Bardeen parameters
$\zeta_n (\rho)$ sufficiently:
\begin{equation}
 \phi_1 (0) = \frac{1}{\kappa} 10, \quad \quad
 \phi_2 (0) = \frac{1}{\kappa} 10^{-l}, 
\end{equation}
where $l$ is a positive number. 
$\alpha(\rho)$ is written by $\alpha(\rho) = 10^k$ where $k$ increases for the time 
evolution $\rho \to 0$.
Since we adopt the approximation that $\rho_2$ ($s_2$) does not govern the cosmic 
energy density $\rho$, we obtain $k < 2l$. 
The non-Gaussianity parameter $f_{NL}$ of the model (\ref{Nmodel1}) is calculated as
\begin{equation}
 f_{NL} = \frac{1}{10^2} \frac{1 + 10^{3k-2l-2}}{(1+10^{2k-2l-2})^2}.
\end{equation}
The exponent evaluation method yields 
\begin{align}
 &k< \frac{2(l+1)}{3} \quad \quad \quad &f_{NL}&=10^{-2},\\
 &\frac{2(l+1)}{3} <k <l+1 \quad \quad \quad &f_{NL}&=10^{3k-2l-4},\\
 & l+1 <k < 2l \quad \quad \quad &f_{NL}&=10^{-k+2l}.
\end{align}
For $2(l+1)/3 <k<l+1$, $f_{NL}$ increases and reaches the maximum $f_{NL}=10^{l-1}$
at $k=l+1$.
For $l+1 <k<2l$, $f_{NL}$ decreases and reaches $f_{NL}=1$ at $k=2l$. 
The non-Gaussianity parameter $f_{NL}$ takes a large value transiently.
So if we want to obtain the large $f_{NL}$ from the present observation, 
we need $\rho_2$ to decay into radiation at $k=l+1$. 
Next we consider the period when the second component $\rho_2$ gets to dominate
the cosmic energy density $\rho$. 
In this period, the first component $\rho_1$ is subdominant.
So we can use (\ref{evowithoutsol4}) and get
\begin{eqnarray}
 N &=& N(1) + \frac{1}{3} \ln{ \left[ \frac{m^2_2 \phi^2_2 (0)}{2 \rho} 
 \left\{ \frac{\kappa^2 \rho_1 (1)}{3 m^2_2} \right\}^{3/4} 
\frac{1}{(1-\kappa^2 \phi^2_2(0) /6)^{3/4}} \right] } \notag\\
&& + \frac{1}{3} \left( \frac{2}{m^2_2 \phi^2_2 (0)} \right)^{4/3}
\frac{3 m^2_2}{\kappa^2} \left(1-\frac{\kappa^2 \phi^2_2(0)}{6} \right)
\rho^{1/3},
\end{eqnarray}
which is simplified into the model (\ref{modulatedN}) for small $\rho$.
Then we can obtain the evaluation of $f_{NL}$
(\ref{modulatedcase1})(\ref{modulatedcase2})(\ref{modulatedcase3}).
If we want $f_{NL}$ of order of unity, we need $l>1$.

There is a case where the contribution to $N$ from the second component
$\rho_2 (1)$ dependent term increases but is bounded in spite of $g_2 < g_1$.
In this case, the non-Gaussianity parameter $f_{NL}$ cannot grow into a
significant value.
We consider the case where $\phi_a$ ($a=1,2$) with mass $m_a$ ($m^2_2 < m^2_1$)
are in the slow rolling phase.
In this case, $g_a = 2 m^2_a /\kappa^2 \rho$. Unlike the previous case,
$g_a$ depends on the cosmic energy density $\rho$.
The ratio of the energy density of the second component $s_2$ and the 
logarithm of the scale factor $N$ are given by
\begin{eqnarray}
 s_2 &=& \frac{\rho_2(0)}{\rho(0)} 
 \left( \frac{\rho}{\rho(0)} \right)^{-1+m^2_2/m^2_1},\\
 N &=& - \frac{\kappa^2}{2 m^2_1} (\rho-\rho(0)) 
 + \frac{\kappa^2}{2 m^2_1} \frac{-m^2_1 +m^2_2}{m^2_2} 
 \rho_2 (0) \left[ \left( \frac{\rho}{\rho(0)}\right)^{m^2_2 / m^2_1} -1 \right]
\end{eqnarray}
where $0$ in $\rho(0)$, $\rho_a (0)$ implies the first horizon crossing time and
$\rho(0)=\rho_1(0)+\rho_2(0)$ where $\rho_a (0) = m^2_a \phi^2_a (0)/2$.
From the above expression of $N$, we can verify that the $\rho_2 (0)$ dependent term 
is bounded and suppressed by $\kappa^2 \phi^2_2 (0)$, therefore $f_{NL}$ is suppressed
by $10^{-2}$.

\subsubsection{the case with the cosmological term}

In cases where the cosmological term $U_0$ exists, we use $\sigma :=\rho -U_0$ as the 
evolution parameter.
In these cases, the evolution equations corresponding to (\ref{evolutionrhoN}),
(\ref{evolutionrhosa}) are the same evolution equations (\ref{evolutionrhoN}),
(\ref{evolutionrhosa}) but all $\rho$'s are replaced with $\sigma$'s.
$s_A$ is defined by $s_A:= \rho_A /\sigma$ and satisfies $\sum_A s_A =1$.
When $|s_2| \ll 1$, linearizing with respect to $s_2$ gives (\ref{evowithouteq1}), 
(\ref{evowithouteq2}) but all $\rho$'s are replaced with $\sigma$'s.
In the same way, when $|s_1| \ll 1$, linearizing with respect to $s_1$ gives 
(\ref{evowithouteq3}), (\ref{evowithouteq4}) but all $\rho$'s are replaced with $\sigma$'s.

We consider two scalar fields $\phi_a$ ($a=1,2$) which move on the potential given by
\begin{equation}
 \rho = U_0 + \sum_{a=1}^2 \rho_a, \quad \quad \quad
 \rho_a = \frac{1}{2} \eta_a \phi^2_a.
\end{equation}
In this case, the $g$ factor of the scalar field $\phi_a$ ($a=1,2$) is given by
$g_a = 2 \eta_a /\kappa^2 \rho \cong 2 \eta_a /\kappa^2 U_0$.
In order that the inflation can solve the horizon problem, we assume that 
$g_a \sim 10^{-2}$.
We assume that $\eta_1 > \eta_2$. 
First the energy of the scalar field $\phi_1$, $\rho_1$ dominate $\sigma$ and next
the energy of the scalar field $\phi_2$, $\rho_2$ grows and gets to dominate $\sigma$.
In the first period where $\rho_1$ dominates $\sigma$, $N$ is given by
\begin{equation}
 N= - \frac{\kappa^2 U_0}{2 \eta_1} \ln{ \frac{\sigma}{\rho_1(0)} }
 + \frac{\kappa^2 U_0}{2 \eta_1} \frac{\rho_2(0)}{\rho_1(0)} 
 \left( \frac{\rho_1(0)}{\sigma} \right)^{1 - \eta_2 /\eta_1},
\label{hybridperiod1} 
\end{equation}
where 
\begin{equation}
 \rho_a (0):= \frac{1}{2} \eta_a \phi^2_a (0).
\end{equation}
In the second period where $\rho_2$ dominates $\sigma$, $N$ is given by
\begin{equation}
 N= - \frac{\kappa^2 U_0}{2 \eta_2} \ln{ \frac{\sigma}{\rho_2(0)} }
 + \frac{\kappa^2 U_0}{2 \eta_2} \frac{\rho_1(0)}{\rho_2(0)} 
 \left( \frac{\rho_2(0)}{\sigma} \right)^{1 - \eta_1 /\eta_2},
\label{hybridperiod2} 
\end{equation}
which can be obtained when all the subscripts $1$, $2$ are exchanged in the 
expression of $N$ in the first period (\ref{hybridperiod1}).

We consider the case that $\eta_1 >0$, $\eta_2 <0$.
In this case, in the first period $\sigma$ is positive, decreases for the 
time evolution, and then the $\rho_2(0)$ dependent term in $N$ 
(\ref{hybridperiod1}) grows boundlessly.
In the first period, $N$ can be simplified into the model given by
\begin{equation}
 N= 10^2 \ln{\phi_1 (0)} - \kappa^2 \alpha(\sigma) \phi^2_2 (0),
\end{equation}
up to the $\sigma$ dependent part which is not related with the Bardeen 
parameters $\zeta_n (\rho)$,
which yields the non-Gaussianity parameter $f_{NL}$ given by
\begin{equation}
 f_{NL} = \left( -10^6 \frac{1}{\phi^4_1 (0)} 
        - \kappa^6 \alpha^3 (\sigma) \phi^2_2 (0) \right) \Big/ 
    \left( 10^4 \frac{1}{\phi^2_1 (0)} 
        + \kappa^4 \alpha^2 (\sigma) \phi^2_2 (0) \right)^2.
\end{equation}
We put 
\begin{equation}
 \phi_1 (0)= \frac{1}{\kappa} 10^m, \quad
 \phi_2 (0)= \frac{1}{\kappa} 10^{-l}, \quad
 \alpha (\sigma) = 10^k,
\end{equation}
where $k$ grows for the time evolution and $k <2l+2$ for the condition
that $\rho_2$ is subdominant.
The non-Gaussianity parameter $f_{NL}$ is written by
\begin{equation}
 f_{NL} = \frac{-10^{6-4m}-10^{3k-2l}}{(10^{4-2m}+10^{2k-2l})^2}.
\end{equation}
The exponent evaluation method gives
\begin{align}
 &k<k_1       \quad \quad \quad &f_{NL}&=-10^{-2},\label{1positiveI1}\\ 
 &k_1 <k <k_2 \quad \quad \quad &f_{NL}&=-10^{3k-2l-8+4m},\label{1positiveI2}\\ 
 &k_2 <k <k_3 \quad \quad \quad &f_{NL}&=-10^{-k+2l},\label{1positiveI3}
\end{align}
where
\begin{equation}
 k_1 := 2 -\frac{4}{3}m + \frac{2}{3}l, \quad
 k_2 := 2 -m + l, \quad
 k_3 := 2 l +2. 
\end{equation}
For $k_1 <k <k_2$, the absolute value of $f_{NL}$ grows and reaches the maximum
$f_{NL}=10^{-2+m+l}$ at $k=k_2$.
For $k_2 <k <k_3$, the absolute value of $f_{NL}$ decreases and reaches  
$f_{NL}=10^{-2}$ at $k=k_3$.
In the second period, $\sigma$ is negative, $- \sigma$ grows, and the $\rho_1(0)$
dependent term in $N$ (\ref{hybridperiod2}) decays.
$N$ in the second period (\ref{hybridperiod2}) can be simplified into the model
given by
\begin{equation}
 N= - 10^2 \ln{\phi_2(0)} + \kappa^2 \beta(\sigma) \phi^2_1 (0),
\end{equation}
up to the $\sigma$ dependent part which is not related with the Bardeen 
parameters $\zeta_n (\rho)$.
Putting 
\begin{equation}
 \phi_1 (0)= \frac{1}{\kappa} 10^m, \quad
 \phi_2 (0)= \frac{1}{\kappa} 10^{-l}, \quad
 \beta (\sigma) = 10^{-k},
\end{equation}
where $k$ increases for the time evolution and satisfies $k >2m-2$ for the condition
that $\rho_1$ is subdominant, we obtain
\begin{eqnarray}
 f_{NL} &=& \left( 10^6 \frac{1}{\phi^4_2 (0)} 
        + \kappa^6 \beta^3 (\sigma) \phi^2_1 (0) \right) \Big/ 
    \left( 10^4 \frac{1}{\phi^2_2 (0)} 
        + \kappa^4 \beta^2 (\sigma) \phi^2_1 (0) \right)^2 \notag\\
  &=& \frac{10^{6+4l} +10^{-3k+2m}}{(10^{2l+4} +10^{-2k+2m})^2} \notag\\
  &\cong& 10^{-2}.      
\end{eqnarray}

We consider the case $\eta_1, \eta_2 <0$ ($\eta_1 > \eta_2$).
In the first period, $\sigma$ is negative, $-\sigma$ increases, and 
the $\rho_2(0)$ dependent term in $N$ (\ref{hybridperiod1}) grows
boundlessly.
In the first period, $N$ (\ref{hybridperiod1}) can be simplified into
the model given by
\begin{equation}
 N= - 10^2 \ln{\phi_1 (0)} - \kappa^2 \alpha(\sigma) \phi^2_2 (0),
\end{equation}
up to the $\sigma$ dependent part which is not related with the Bardeen 
parameters $\zeta_n (\rho)$,
which yields the non-Gaussianity parameter $f_{NL}$ given by
\begin{eqnarray}
 f_{NL} &=& \left( 10^6 \frac{1}{\phi^4_1 (0)} 
        - \kappa^6 \alpha^3 (\sigma) \phi^2_2 (0) \right) \Big/ 
    \left( 10^4 \frac{1}{\phi^2_1 (0)} 
        + \kappa^4 \alpha^2 (\sigma) \phi^2_2 (0) \right)^2 \notag\\
&=& \frac{10^{6+4m} -10^{3k-2l}}{(10^{2m+4} +10^{2k-2l})^2},
\end{eqnarray}
by putting 
\begin{equation}
 \phi_1 (0)= \frac{1}{\kappa} 10^{-m}, \quad
 \phi_2 (0)= \frac{1}{\kappa} 10^{-l}, \quad
 \alpha (\sigma) = 10^k,
\end{equation}
where in the first period $\rho_1$ is dominant $m<l$ and
$k$ increases for the time evolution, $k<2l+2$ from the 
condition that $\rho_2$ is subdominant.
The exponent evaluation method gives
\begin{align}
 &k<k_1       \quad \quad \quad &f_{NL}&=10^{-2},\label{1negativeI1}\\ 
 &k_1 <k <k_2 \quad \quad \quad &f_{NL}&=-10^{3k-2l-4m-8},\label{1negativeI2}\\ 
 &k_2 <k <k_3 \quad \quad \quad &f_{NL}&=-10^{-k+2l},\label{1negativeI3}
\end{align}
where
\begin{equation}
 k_1 := \frac{2}{3}l + \frac{4}{3}m+2, \quad
 k_2 := l +m+2, \quad
 k_3 := 2 l +2. 
\end{equation}
For $k_1 <k <k_2$, the absolute value of $f_{NL}$ grows, reaches the maximum
$f_{NL} = -10^{l-m-2}$ at $k=k_2$.
For $k_2 <k <k_3$, the absolute value of $f_{NL}$ decreases, reaches 
$f_{NL} = -10^{-2}$ at $k=k_3$.
In the second period, $\sigma$ is negative, $- \sigma$ grows and the 
$\rho_1(0)$ dependent term in $N$ (\ref{hybridperiod2}) decays.
$N$ in the second period (\ref{hybridperiod2}) can be simplified into
the model given by
\begin{equation}
 N= - 10^2 \ln{\phi_2 (0)} - \kappa^2 \beta(\sigma) \phi^2_1 (0),
\end{equation}
up to the $\sigma$ dependent part which is not related with the Bardeen 
parameters $\zeta_n (\rho)$, which yields
\begin{eqnarray}
 f_{NL} &=& \left( 10^6 \frac{1}{\phi^4_2 (0)} 
        - \kappa^6 \beta^3 (\sigma) \phi^2_1 (0) \right) \Big/ 
    \left( 10^4 \frac{1}{\phi^2_2 (0)} 
        + \kappa^4 \beta^2 (\sigma) \phi^2_1 (0) \right)^2 \notag\\
  &=& \frac{10^{6+4l} -10^{-3k-2m}}{(10^{2l+4} +10^{-2k-2m})^2} \notag\\
  &\cong& 10^{-2},      
\end{eqnarray}
putting
\begin{equation}
 \phi_1 (0)= \frac{1}{\kappa} 10^{-m}, \quad
 \phi_2 (0)= \frac{1}{\kappa} 10^{-l}, \quad
 \beta (\sigma) = 10^{-k},
\end{equation}
where $k$ grows for the time evolution, and $k >-2m-2$ from
the condition that $\rho_1$ is subdominant.
The above results about two scalar fields in the vacuum domination can be
investigated by the method of the $\tau$ function. \cite{Hamazaki2008.2} 
The same results given in this subsection are reproduced using the $\tau$
function in the Appendix $C$.

The mechanism which produces the large $f_{NL}$ depends on the fact that
$g_1 > g_2$ and that $s_1$ is dominant in the first time and that $s_2$
begins to govern the cosmic energy density $\rho$ gradually. 
So when $s_1$ is the radiation $g_1 =4$, and $s_2$ is the scalar field with 
the negative mass $g_2 = 2 \eta / \kappa^2 U_0$ ($\eta < 0$),
the non-Gaussianity $f_{NL}$ can grow transiently, because the third term 
which depends on $\rho_2 (0)$  
\begin{equation}
 N= N(0)- \frac{1}{4} \ln{ \frac{\sigma}{\rho_1(0)} }
 + \frac{1}{4} \frac{\rho_2(0)}{\rho_1(0)} 
 \left( \frac{\rho_1(0)}{\sigma} \right)^{1 - \eta /2 \kappa^2 U_0}, 
\end{equation}
can grow boundlessly.

\section{Discussion}

In this paper, the first half, that is sections $2$, $3$ is devoted to the general 
considerations about the gauge invariant nonlinear cosmological perturbation theory
on superhorizon scales and the latter half, that is section $4$ is devoted to the 
investigations of the concrete physical systems.
Here we discuss how the general theory in the first half is used in the analysis 
of the concrete physical systems in the latter half.

($A$) In subsection $2.1$, we gave the definitions of all the types of gauge invariant
perturbation variables.
The numerical data obtained by the cosmological observations are related with the 
gauge invariant perturbation variables, since both quantities do not depend on how 
we set up the spacetime coordinate system.
Therefore it is desirable to express all the physical laws in the form closed by
the gauge invariant quantities only.
By the subsection $2.1$, it is guaranteed that the perturbation variables given
by operating $D(\rho)$ derivatives on the scalar like objects such as $\ln{a}$,
$s_A := \rho_A / \rho$ are gauge invariant.
In the analysis of the concrete physical systems in the latter half of the present 
paper, the adiabatic perturbation variable $D(\rho)^n \ln{a}$, the entropic 
perturbation variables $D(\rho)^n s_A$ are used as the independent variables and
the total energy density $\rho$ is used as the evolution parameter.
This formalism by the A/E decomposition given in the subsection $3.2$ and
the $\rho$ philosophy proposed in the subsection $3.3$ is very useful for the physical 
interpretations of the results obtained in the latter half of the paper.

($B$) In the subsection $2.2$, we construct the metric junction formalism as the 
method treating the sudden change of the equation of state.
The metric junction formalism given in the subsection $2.2$ is used in the analysis
of the concrete physical systems in the latter half of the paper,
since they contain the transitions such as the slow rolling oscillatory transition 
and the reheating transition.
Since it is proven in the subsection $2.2$ that our A/E perturbation variables
$D(\rho)^n \ln{a}$, $D(\rho)^n s_A$ are continuous across the matching surface 
defined by $\rho = {\rm const}$, this set of variables must be useful in the research
of the transitions even if we assume that the transition do not occur instantly.

($C$) Based on the $\rho$ philosophy proposed in the subsection $3.3$, in section $4$
the evolutions of $\ln{a}$, $s_A$ are described as the functions of the evolution 
parameter $\rho$.
The evolutions of our A/E perturbation variables $D(\rho)^n \ln{a}$, $D(\rho)^n s_A$
are given by operating $D(\rho)$ derivatives on the solutions of $\ln{a}$, $s_A$
expressed as the functions of $\rho$.
Since $D(\rho) f(\rho) =0$ for an arbitrary function of $\rho$; $f(\rho)$, the 
$\rho$ dependences of $\ln{a}$, $s_A$ are directly reflected to the $\rho$ dependences
of $D(\rho)^n \ln{a}$, $D(\rho)^n s_A$.
For this reason, by the formalism prepared in section $3$, in section $4$ we can manifestly
clarify the time evolutions in which the entropic perturbation of the energetically 
subdominant component makes the adiabatic perturbation $D(\rho)^n \ln{a}$ grow.
We can show that $D(\rho)^n \ln{a}$, the non-Gaussanity $f_{NL}$ grow considerably
and transiently when the energy ratio $s_2$ of the energetically subdominant component
which has the $g$ factor smaller than the $g$ factor of the dominant component $\rho_1$  
and which is supported by the extremely small scalar field expectation value, begins
to increase.
Since until now as for $D(\rho)^n \ln{a}$ only expression of the final state when 
the growth of $D(\rho)^n \ln{a}$ has already ended, has been given,
the $\rho$ philosophy given in section $3$ is superior in that the time evolution and 
the mechanism of the growth of the adiabatic perturbation variable called the Bardeen 
parameters $D(\rho)^n \ln{a}$, the non-Gaussianity $f_{NL}$ can be described.

In the following paragraphs, the two points which was not treated in the first 
section are discussed.
The first point is on the existence of more than two sources of cosmological 
perturbations.
The second point is on the modeling of the abrupt cosmic evolution by the 
metric junction across the spacelike hypersurface.

($1$)The present amplitude of the Bardeen parameter $\zeta_n (\rho)$ can be decomposed 
as $\zeta_n (\rho) = \zeta_{n \> {\rm sl}} (\rho)+\zeta_{n \> {\rm ent}} (\rho)$.
$\zeta_{n \> {\rm sl}} (\rho)$ is the component generated from the adiabatic mode
in the slow rolling phase, and $\zeta_{n \> {\rm ent}} (\rho)$ is the adiabatic 
component generated from the entropy mode in the slow rolling phase by successive 
universe evolution.
In many excellent papers, many authors insisted that a significant large nonlinearity 
can be generated in the inhomogeneous end of the inflation \cite{Lyth2005.2} \cite{Sasaki2008},
in the modulated reheating \cite{Dvali2004}, in the curvaton scenario \cite{Ichikawa2008}
and in the vacuum dominated inflation \cite{Alabidi2006} \cite{Byrnes2009}.
Unfortunately in the partial studies, without any plausible reasons it is assumed that
$\zeta_{n \> {\rm sl}} (\rho)$ is negligibly small compared with
$\zeta_{n \> {\rm ent}} (\rho)$ and the non-Gaussianity parameter $f_{NL}$ is calculated
from $\zeta_{n \> {\rm ent}} (\rho)$ only.
However the inflaton which drives the universe expansion enough to solve the horizon 
problem, the flatness problem, but do not generate any cosmological perturbations,
does not exist.
In this point of view, it is wonderful that the authors of the paper \cite{Ichikawa2008}
tried to treat the contribution of the inflaton $\zeta_{n \> {\rm sl}} (\rho)$ 
and the contribution of the curvaton $\zeta_{n \> {\rm ent}} (\rho)$
with equal importance from the standpoint of the mixed scenario.
In this paper, we investigate whether a significant non-Gaussianity $f_{NL}$
is generated in the successive universe evolution taking into account the cosmological 
perturbations generated in the slow rolling phase.
When we analyze the systems where more than two factors are concerned, 
the exponent evaluation method presented in this paper is very efficient.

($2$)In the early universe, there exists a period before which and after which the 
dynamical behaviors of each component are very different.
As for the scalar field with mass $m$, while $m \ll H$, where $H$ is the Hubble parameter,
holds, the scalar field is in the slow rolling phase when its energy density changes mildly
compared with the cosmic expansion, and while $H \ll m$ holds, the scalar field is in the 
oscillatory phase when its energy density behaves like a dust fluid.\cite{Kodama1996}
\cite{Hamazaki2002} \cite{Hamazaki2004} \cite{Hamazaki2008}
In the reheating \cite{Hamazaki1996} \cite{Hamazaki2008}, the energy of the oscillatory scalar 
field behaving like a dust fluid is transformed into that of the radiation fluid.
In the hybrid inflation, the energy of the slow rolling scalar field is transformed into that 
of the oscillatory scalar field and into that of the radiation fluid on the bifurcation set.
Such phase transitions are quite complicated and the completely rigorous mathematical treatment
is beyond our scope. 
For example, we consider the slow rolling-oscillatory transition. 
In the $m \ll H$ region and in the $H \ll m$ region, the expansion schemes investigating 
the dynamical behaviors of the scalar field can be developed 
with $m/H$ and $H/m$ as the expansion parameter, respectively.\cite{Hamazaki2008.2}
\cite{Kodama1996} \cite{Hamazaki2002} \cite{Hamazaki2004} \cite{Hamazaki2008}
But at the transition period $H \sim m$, any expansion schemes cannot be developed
because of no expansion parameters.
However, in spite of complicated behaviors at the transition, the period of the transition
can be thought to be short compared with the periods before and after the transition characterized
by $m \ll H$, $H \ll m$, respectively.
Therefore we think the transitions as the instantaneous transient phenomena and may treat such 
transitions as the metric junctions across the spacelike hypersurfaces. 
In the above reasons, in our paper, the metric junction formulation on the cosmological 
perturbations in the long wavelength limit, linear and nonlinear, are constructed. 
On the spacelike hypersurface defined by $H=m$, the spacetime governed by the slow rolling scalar field
and the spacetime governed by the oscillatory scalar field is connected. 
In case of reheating, On the spacelike hypersurface defined by $H=\Gamma$ where $\Gamma$ is the decay 
constant of the scalar field, the spacetime governed by the oscillatory scalar field
and the spacetime governed by the radiation fluid is connected.

\section*{Acknowledgments}

The author would like to thank Professors H. Kodama, 
J. Yokoyama \cite{Takahashi2009} and M. Sasaki \cite{Takamizu2010}
for continuous encouragements.
He would like to thank Professor Mukohyama for writing his educational
paper \cite{Mukohyama2000}, from which he learned a lot about the 
metric junction.

\appendix

%koko

%% Equation numbering %%
% koko de Appendix no siki banngou wo (A,1) ni suru

\catcode`\@=11

\@addtoreset{equation}{section}   % Makes \section reset 'equation' counter.
\def\theequation{\Alph{section}.\arabic{equation}}

%koko

\section{Proof of Proposition $8$}

Solving the matching condition $C(t, {\bm x}, \lambda)=0$ with respect
to $t$ gives $t=t({\bm x}, \lambda)$.
By differentiating $C(t, {\bm x}, \lambda)=0$ with respect to $\lambda$,
$x^k$, we obtain
\begin{equation}
 \frac{d t}{d \lambda} = - \frac{1}{\dot{C}} \frac{d C}{d \lambda},
 \quad \quad
 \frac{\partial t}{\partial x^k} = - \frac{1}{\dot{C}} \partial_k C.
\end{equation}
Differentiating $[S]^+_- =0$ with respect to $\lambda$, $x^k$ gives
\begin{equation}
 \left[ 
 \frac{d S}{d \lambda} + \dot{S} \frac{d t}{d \lambda}
 \right]^+_- =0, \quad \quad
 \left[ \partial_i S + \dot{S} \frac{\partial t}{\partial x^i}
 \right]^+_- =0
\end{equation}
From the two sets of equations, we obtain $[D(C) S]^+_- =0$, 
$[D_i (C) S]^+_- =0$.

\section{the junction condition in the linear perturbation theory}

In this section, we consider the metric junction across the matching
hypersurface characterized by $\tilde{C} = 0$ where $\tilde{C}$ is the scalar
quantity in the linear perturbation theory, since the previous papers
treating this problem \cite{Deruelle1995} \cite{Martin1998} 
\cite{Copeland2007} sometimes contain some typographical errors, 
derive the matching conditions without keeping the gauge invariance 
completely and are lacking in the physical interpretations
from the viewpoint of the long wavelength limit.
The notaions used in this section are based on the papers. 
\cite{Kodama1984} \cite{Kodama1998} \cite{Hamazaki2008}
The metric tensor is given by
\begin{eqnarray}
 \tilde{g}_{00} &=& - (1 + 2 A Y),\\
 \tilde{g}_{0i} &=& - a B Y_i,\\
 \tilde{g}_{ij} &=& a^2 [(1 + 2 H_L Y) \delta_{ij} + 2 H_T Y_{ij}],
\end{eqnarray}
where $Y$, $Y_i$ and $Y_{ij}$ are harmonic scalar, vector and tensor
for a scalar perturbation with wavenumber ${\bm k}$:
\begin{equation}
 Y:= e^{i {\bm k}{\bm x}}, \quad Y_i:= - i \frac{k_i}{k} Y, \quad
 Y_{ij}:= \left( \frac{1}{3} \delta_{ij} - \frac{k_i k_j}{k^2} 
          \right) Y,
\end{equation}
where $k^2 := \sum_i k_i k_i$.
The energy momentum tensor of the total system is given by
\begin{equation}
 \tilde{T}_{\mu \nu} = (\tilde{\rho} + \tilde{P}) 
 \tilde{u}_{\mu} \tilde{u}_{\nu} + \tilde{P} \tilde{g}_{\mu \nu},
\end{equation}
where $\tilde{\rho}$, $\tilde{P}$ and $\tilde{u}_{\mu}$ are the energy
density, the pressure, and the four velocity of the total system.
For the scalar quantities $\tilde{S}=(\tilde{\rho}, \tilde{P})$,
$\tilde{S}$ is expanded as $\tilde{S} = S + \delta S Y$, and the 
four velocity $\tilde{u}_{\mu}$ is written by
\begin{eqnarray}
 \tilde{u}_0 &=& -(1+A Y),\\ 
 \tilde{u}_i &=& a (v - B) Y_i.
\end{eqnarray}
We define the gauge dependent geometrical quantities as 
\begin{equation}
 {\cal R} := H_L + \frac{1}{3} H_T, \quad \quad
 \sigma_g := \frac{a}{k} \dot{H}_T - B. 
\end{equation}
For the scalar quantity $\tilde{S}$, 
\begin{equation}
 D S := \delta S - \frac{\dot{S}}{H} {\cal R}
\end{equation}
is gauge invariant.
For the four velocity $\tilde{u}_{\mu}$, the variable defined by
\begin{equation}
 Z:= {\cal R} - \frac{a H}{k} (v - B)
\end{equation}
is gauge invariant.
The Newtonian potential $\Phi$ defined by
\begin{equation}
 \Phi:= {\cal R} - \frac{a H}{k} \sigma_g
\end{equation}
is gauge invariant.

We consider the metric junction across the hypersurface $\Sigma$
defined by
\begin{equation}
 x^0 = t_{\times} + \delta Z^0 Y, \quad \quad
 x^i = y^i + \delta Z Y^i,
\end{equation}
which connects the future spacetime ${\cal M}_+$ and the past
spacetime ${\cal M}_-$. 
From $\delta Z^0$, $\delta Z$, we can define two gauge invariant
quantities as
\begin{eqnarray}
 \phi^0 &:=& \delta Z^0 + \frac{1}{H} {\cal R},\\
 \phi &:=& \delta Z - \frac{1}{k} H_T.
\end{eqnarray}
We consider the case where the matching hypersurface $\Sigma$ is defined 
by $\tilde{C}=0$.
In this case
\begin{equation}
 \phi^0 = - \frac{1}{\dot{C}} D C.
\end{equation}
The normal vector of the matching hypersurface which points from
${\cal M}_-$ to ${\cal M}_+$ is given by
\begin{equation}
 \tilde{n}_{\mu} = - {\rm sgn}(\dot{\tilde{T}}) 
 [ - \tilde{g}^{\rho \sigma} 
 \partial_{\rho} \tilde{T} \partial_{\sigma} \tilde{T}]^{-1/2}
 \partial_{\mu} \tilde{T}.
\end{equation}
We define the intrinsic metric, the extrinsic curvature, the intrinsic
energy momentum tensor by
\begin{eqnarray}
 \tilde{q}_{ij} &:=& \tilde{e}^{\mu}_i \tilde{e}^{\nu}_j
 (\tilde{g}_{\mu \nu} + \tilde{n}_{\mu} \tilde{n}_{\nu}),\\
 \tilde{K}_{ij} &:=& \tilde{e}^{\mu}_i \tilde{e}^{\nu}_j
 \tilde{\nabla}_{\mu} \tilde{n}_{\nu},\\
 \tilde{T}_{nn} &:=& \tilde{n}^{\mu} \tilde{n}^{\nu}
 \tilde{T}_{\mu \nu},\\
 \tilde{T}_{ni} &:=& \tilde{n}^{\mu} \tilde{e}^{\nu}_i 
 \tilde{T}_{\mu \nu}, 
\end{eqnarray}
where $\tilde{e}^{\mu}_i := \partial x^{\mu} / \partial y^i$.
These quantities of the matching hypersurface $\Sigma$ defined by
$\tilde{C} = 0$ can be written as

\begin{eqnarray}
 \tilde{q}_{ij} &=& a^2 \left\{ \delta_{ij} + 2 \delta_{ij} Y
\left( - H \frac{D C}{\dot{C}} + \frac{k}{3} \phi
\right) + 2 Y_{ij} (- k \phi)
\right\},\\
 \tilde{K}_{ij} &=& a^2 H \delta_{ij} \nonumber\\
&& + \left\{ \left( - 2 a^2 H^2 - a^2 \dot{H} + \frac{k^2}{3} \right)
 \frac{D C}{\dot{C}} + \frac{2}{3} a^2 H k \phi
 + \frac{a^2}{2} H \frac{D \rho}{\rho}
 \right\} Y \delta_{ij}\nonumber\\
 && + \left( 
 - 2 a^2 H k \phi - k^2 \frac{D C}{\dot{C}} - \frac{k^2}{H} \Phi
      \right) Y_{ij}, \\
 \tilde{T}_{nn} &=& \rho + 
 \left( - \frac{\dot{\rho}}{\dot{C}} D C + D \rho
 \right) Y,\\
 \tilde{T}_{ni} &=& (\rho + P) \left( k \frac{D C}{\dot{C}} 
 + \frac{k}{H} Z \right),     
\end{eqnarray}
where right hand sides are written in the gauge invariant form.
The metric junction conditions across the matching hypersurface $\Sigma$
are given by
\begin{equation}
 [\tilde{q}_{ij}]^+_- = [\tilde{K}_{ij}]^+_- = [\tilde{T}_{nn}]^+_-
 = [\tilde{T}_{ni}]^+_- = 0,
\end{equation}
which yield the matching conditions in the long wavelength limit:
in the background level,
\begin{equation}
 [a]^+_- = [H]^+_- = [\rho]^+_- =0,
\end{equation}
and in the perturbation level,
\begin{equation}
 [\phi]^+_- = 0, 
 \label{metricjunctionlinear1}
\end{equation}
\begin{equation}
 [k^2 \Phi]^+_- = 0, 
 \label{metricjunctionlinear2}
\end{equation}
\begin{equation}
 \left[ \frac{D C}{\dot{C}} \right]^+_- = 
 \left[ - \frac{\dot{\rho}}{\dot{C}} D C + D \rho \right]^+_- =0, 
 \label{metricjunctionlinear3}
\end{equation}
\begin{equation}
 \left[ (\rho + P) \left( \frac{D C}{\dot{C}} 
 + \frac{1}{H} Z \right) \right]^+_- = 0.
 \label{metricjunctionlinear4}
\end{equation}
Owing to our previous paper \cite{Hamazaki2008}, in the long wavelength
limit, the solution of $D S$ where $S$ is the scalar quantity is given by
\begin{equation}
 D S = D S^{\sharp} + \frac{\dot{S}}{H} c \int_{t_0} dt \frac{1}{a^3},
\end{equation}
where

\begin{equation}
 D S^{\sharp} := \frac{\partial S}{\partial C_{\star}} -
 \frac{\dot{S}}{\dot{a}} \frac{\partial a}{\partial C_{\star}}
\end{equation}
where $\partial S / \partial C_{\star}$, $\partial a / \partial C_{\star}$
are the derivatives of the background quantities $S$, $a$ with respect to
the solution constant $C_{\star}$ and $c$ is a constant characterizing the
adiabatic decaying mode.
The solution of the Newtonian potential $\Phi$ is given by
\begin{equation}
 k^2 \Phi = \frac{3 H}{a} c + O(k^2).
\end{equation}
Therefore (\ref{metricjunctionlinear2}), (\ref{metricjunctionlinear3}) 
give
\begin{eqnarray}
 && [c]^+_- =0,
 \label{junctionlinearinterpret1}\\
 && [D(C) a]^+_- = [D(C) \rho]^+_- =0,
 \label{junctionlinearinterpret2}
\end{eqnarray}
where as for scalar like object $S$
\begin{equation}
 D(C) S := \frac{\partial S}{\partial C_{\star}} 
- \frac{\dot{S}}{\dot{C}} \frac{\partial C}{\partial C_{\star}}.
\end{equation}
(\ref{junctionlinearinterpret1}), (\ref{junctionlinearinterpret2})
are consistent with the metric junction conditions of the full 
nonlinear gradient expansion case represented by (\ref{nonlinearadiabaticdecaying}) 
and the Proposition $8$.

\section{the analyses of the evolution of the multiple vacuum dominated scalar fields
by the $\tau$ function}

The $\tau$ function was presented as the method of analyzing the evolution of the multiple
scalar fields. \cite{Hamazaki2008.2}
Under the slow rolling approximation, the evolution of the scalar fields is described by
\begin{equation}
 \frac{d \phi_a}{d N} = - \frac{1}{\kappa^2 U} \frac{\partial U}{\partial \phi_a},
 \label{originalscalar}
\end{equation}
which are decomposed as
\begin{equation}
 \frac{d \phi_a}{d \tau} = -\frac{\partial U}{\partial \phi_a}, \quad \quad
 \frac{d N}{d \tau} = \kappa^2 U,
 \label{taufunction}
\end{equation}
introducing the $\tau$ function as the new evolution parameter.\cite{Hamazaki2008.2}  
It is much easier to treat the new evolution equations (\ref{taufunction}) than
to treat the original evolution equations (\ref{originalscalar}) for many cases.

We consider the vacuum dominated case given by
\begin{equation}
 U = U_0 + \sum_a \frac{1}{2} \eta_a \phi^2_a.
\end{equation}
In this case, the evolution of $\phi_a$ is given by
\begin{equation}
 \phi_a =\phi_a (0) \exp{(-\eta_a \tau)}.
\end{equation}
By using the $\tau$ function as the evolution parameter, the Bardeen parameter
$\zeta_n (\rho)$ are given by
\begin{equation}
 \zeta_n (\rho) = \left( \frac{\partial}{\partial \lambda} 
 - \frac{U_{\lambda}}{U_{\tau}} \frac{\partial}{\partial \tau} \right)^n 
 \left( \kappa^2 \int_0 d \tau U \right),
\end{equation}
where the subscripts $\lambda$, $\tau$ are interpreted as the derivatives with 
respect to $\lambda$, $\tau$, respectively; for example
\begin{equation}
 U_{\lambda \tau}:= 
 \frac{\partial}{\partial \lambda} \frac{\partial}{\partial \tau} U.
\end{equation}
Concretely $\zeta_1 (\rho)$, $\zeta_2 (\rho)$ are given by
\begin{eqnarray}
 \frac{1}{\kappa^2} \zeta_1 (\rho) &=& 
 \int_0 d \tau U_{\lambda} - \frac{U_{\lambda}}{U_{\tau}} U,\\ 
 \frac{1}{\kappa^2} \zeta_2 (\rho) &=& \int_0 d \tau U_{\lambda \lambda} 
 - \frac{U^2_{\lambda}}{U_{\tau}}
 +\frac{U}{U^2_{\tau}} \left( - U_{\lambda \lambda} U_{\tau}+2 U_{\lambda} U_{\lambda \tau}
 - \frac{U^2_{\lambda}}{U_{\tau}} U_{\tau \tau}
 \right),
\end{eqnarray}
and the coefficients $N_a$, $N_{ab}$ are given by the expansions
\begin{equation}
 \zeta_1 (\rho) = \sum_a N_a \frac{d \phi_a (0)}{d \lambda}, \quad \quad
 \zeta_2 (\rho) = \sum_{ab} N_{ab} \frac{d \phi_a (0)}{d \lambda} 
 \frac{d \phi_b (0)}{d \lambda},
\end{equation}
where we assume that all the nonlinear perturbations at the first horizon crossing in the 
inflationary expansion are vanishing:
$d^n \phi_a (0)/ d \lambda^n =0$ ($n \ge 2$).
By using $A(n, k)$ defined by
\begin{equation}
 A(n, k):= \sum_a \eta^n_a \phi^2_a (0) \exp{(- 2 k \eta_a \tau)},
\end{equation}
and collecting the leading order terms with respect to $U_0$, we obtain 
\begin{equation}
 f_{NL} = \frac{1}{\kappa^2 U_0} \left[ 
 \frac{A(2,1) A(3,3)}{A(2,2)^2} - 4 \frac{A(3,2)}{A(2,2)}
 + 2 \frac{A(3,1)}{A(2,1)}
\right].
\label{taunonGauss}
\end{equation}
We consider the system where two scalar fields evolve.
Since only the first term in (\ref{taunonGauss}) can change the exponent for the
moving $\tau$, we concentrate on the first term written by $(f_{NL})_1$ from 
now on.
$(f_{NL})_1$ can be written by
\begin{equation}
 (f_{NL})_1 = \frac{\eta_1}{\kappa^2 U_0} 
 \frac{(1+ \eta^2_r \phi^2_r e_r)(1+ \eta^3_r \phi^2_r e^3_r)}
 {(1+ \eta^2_r \phi^2_r e^2_r)^2},
\end{equation}
where
\begin{equation}
 \eta_r := \frac{\eta_2}{\eta_1}, \quad
 \phi_r := \frac{\phi_2(0)}{\phi_1(0)}, \quad
 e_r := \exp{\{ -2 (\eta_2-\eta_1) \tau \}}.
\end{equation}
In order that the inflation can solve the horizon problem, we assume 
$\eta_a/\kappa^2 U_0 \sim 10^{-2}$.

First we consider the case $\eta_1 >0$, $\eta_2 <0$.
we put
\begin{equation}
 \phi_1 (0) = \frac{1}{\kappa} 10^m, \quad
 \phi_2 (0) = \frac{1}{\kappa} 10^{-l}, \quad
 e_r = 10^p,
\end{equation}
then we get
\begin{equation}
 (f_{NL})_1 = 10^{-2}
 \frac{(1+ 10^{-2(l+m)+p})(1- 10^{-2(l+m)+3p})}
 {(1+ 10^{-2(l+m)+2p})^2}.
\end{equation}
The exponent evaluation method gives
\begin{align}
 &p <p_1 \quad \quad \quad &(f_{NL})_1 =& 10^{-2},\\
 &p_1 <p <p_2 \quad \quad \quad &(f_{NL})_1 =& -10^{3p-2(l+m)-2},\\
 &p_2 <p <p_3 \quad \quad \quad &(f_{NL})_1 =& -10^{-p+2(l+m)-2},\\
 &p_3 <p \quad \quad \quad &(f_{NL})_1 =& -10^{-2},
\end{align}
where
\begin{equation}
 p_1 := \frac{2}{3} (l+m),\quad \quad
 p_2 := (l+m), \quad \quad
 p_3 := 2(l+m).
\end{equation}
The above evaluations agree completely with (\ref{1positiveI1}), 
(\ref{1positiveI2}), (\ref{1positiveI3}) by taking the 
correspondence $p=k+2m-2$.

Next we consider the case where $\eta_a <0$, $\eta_1 > \eta_2$.
We put
\begin{equation}
 \phi_1 (0) = \frac{1}{\kappa} 10^{-m}, \quad
 \phi_2 (0) = \frac{1}{\kappa} 10^{-l}, \quad
 e_r = 10^p,
\end{equation}
where $m<l$, 
then we obtain
\begin{equation}
 (f_{NL})_1 = - 10^{-2}
 \frac{(1+ 10^{-2(l-m)+p})(1+ 10^{-2(l-m)+3p})}
 {(1+ 10^{-2(l-m)+2p})^2}.
\end{equation}
The exponent evaluation method gives
\begin{align}
 &p <p_1 \quad \quad \quad &(f_{NL})_1 =& -10^{-2},\\
 &p_1 <p <p_2 \quad \quad \quad &(f_{NL})_1 =& -10^{3p-2(l-m)-2},\\
 &p_2 <p <p_3 \quad \quad \quad &(f_{NL})_1 =& -10^{-p+2(l-m)-2},\\
 &p_3 <p \quad \quad \quad &(f_{NL})_1 =& -10^{-2},
\end{align}
where
\begin{equation}
 p_1 := \frac{2}{3} (l-m),\quad \quad
 p_2 := (l-m), \quad \quad
 p_3 := 2(l-m).
\end{equation}
The above evaluations agree completely with (\ref{1negativeI1}), 
(\ref{1negativeI2}), (\ref{1negativeI3}) by taking the 
correspondence $p=k-2m-2$.

\addtolength{\baselineskip}{-3mm}

%T1>Biblipgraphy

\end{document}